\documentclass[fleqn,usenatbib]{mnras}

\usepackage[T1]{fontenc}
\usepackage{ae,aecompl}

\usepackage{graphicx}	
\usepackage{amsmath}	
\usepackage{amssymb}	
\usepackage{soul}
\usepackage[normalem]{ulem}
\usepackage[dvipsnames]{xcolor}
\usepackage[normalem]{ulem}
\usepackage{cancel}
\usepackage{lineno}
\linenumbers

\newcommand{\Msun}{\,M$_\odot$}
\newcommand{\rv}{R$_{200}$}
\newcommand{\vdBt}{vdB20}

\usepackage{newtxtext,newtxmath}

\title[The infall region of z$\sim$1 clusters]{Satellite quenching was not important for z$\sim$1 clusters: most quenching occurred during infall.}

\author[S. V. Werner et al.]{
S. V. Werner$^{1}$\thanks{E-mail: stephane.werner@nottingham.ac.uk},
N.\,A.~Hatch$^{1}$, A. Muzzin$^{2}$, R. F.\,J.~van der Burg$^{3}$,  \newauthor
M. L. Balogh$^{4,5}$, G. Rudnick$^{6}$ \& G. Wilson$^{7}$\\
$^{1}$School of Physics and Astronomy, University of Nottingham, Nottingham, NG7 2RD, UK\\
$^{2}$Department of Physics and Astronomy, York University, 4700, Keele Street, Toronto, Ontario, ON MJ3 1P3, Canada\\
$^{3}$European Southern Observatory, Karl-Schwarzschild-Str. 2, 857748, Garching, Germany \\
$^{4}$Department of Physics and Astronomy, University of Waterloo, Waterloo, ON N2L 3G1, Canada\\
$^{5}$Waterloo Centre for Astrophysics, University of Waterloo, Waterloo, ON N2L 3G1, Canada\\
$^{6}$Department of Physics and Astronomy, The University of Kansas, Malott Room 1082, 1251 Wescoe Hall Drive, Lawrence, KS 66045, USA\\
$^{7}$Department of Physics and Astronomy, University of California, Riverside, 900 University Avenue, Riverside, CA 92521, USA
}

\date{Accepted XXX. Received YYY; in original form ZZZ}

\pubyear{2021}

\begin{document}
\nolinenumbers
\label{firstpage}
\pagerange{\pageref{firstpage}--\pageref{lastpage}}
\maketitle

\begin{abstract}
We quantify the relative importance of environmental quenching versus pre-processing in $z\sim1$ clusters by analysing the infalling galaxy population in the outskirts of 15 galaxy clusters at $0.8<z<1.4$ drawn from the GOGREEN and GCLASS surveys. 
We find significant differences between the infalling galaxies and a control sample; in particular, an excess of massive quiescent galaxies in the infalling region. These massive infalling galaxies likely reside in larger dark matter haloes than similar-mass control galaxies because they have twice as many satellite galaxies. Furthermore, these satellite galaxies are distributed in an NFW profile with a larger scale radius compared to the satellites of the control galaxies. Based on these findings, we conclude that it may not be appropriate to use \lq field\rq\ galaxies as a substitute for infalling pre-cluster galaxies when calculating the efficiency and mass dependency of environmental quenching in high redshift clusters. By comparing the quiescent fraction of infalling galaxies at $1<R/$\rv $<3$ to the cluster sample ($R/$\rv $<1$) we find that almost all quiescent galaxies with masses $>10^{11}$\Msun\ were quenched prior to infall, whilst up to half of lower mass galaxies were environmentally quenched after passing the virial radius. This means most of the massive quiescent galaxies in $z\sim1$ clusters were self-quenched or pre-processed prior to infall.

\end{abstract}

\begin{keywords}
Galaxies: clusters: general -- Galaxies: mass function -- Galaxies: evolution -- Galaxies: photometry 
\end{keywords}



\section{Introduction}

The presence of galaxy clusters at $z>1.5$ that host quiescent galaxies with old stellar populations \citep[e.g.][]{Rudnick_2012, Newman_2014, Cook_2016, Nantais2016, Strazzullo2019} poses a challenge to our understanding of environmental quenching. It has long been recognised that galaxy colours \citep{Balogh2004}, ages \citep{Cooper2010}, morphologies \citep{Dressler1980} and star-formation rates (SFRs) \citep{Gomez2003} correlate with environment. Environmental quenching is an all-encompassing term used to describe any process that can cause these correlations by quenching star formation in a manner whose efficiency scales with galaxy density \citep{Peng2010}. Studies of local clusters have revealed several processes that may be responsible for this quenching:  gas starvation, strangulation, ram-pressure, harassment, mergers and tidal stripping \citep{Gunn1972,Dressler1997, Moore1996}.

Clusters of galaxies, as hosts to hundreds of satellite galaxies, are the best places to investigate environmental quenching of satellite galaxies. An estimate of the timescale for satellite quenching comes from the SFRs of star-forming galaxies within clusters: cluster galaxies obey a similar SFR -- stellar mass relation as non-cluster galaxies \citep[][]{Muzzin2012gclass, Old2021}, therefore environmental quenching likely follows a delayed-then-rapid quenching timescale \citep[][]{Wetzel2013}. In this theory cluster galaxies continue to form stars for a ‘delay’ time after falling into the cluster, but then quench so rapidly that few galaxies are observed in the quenching phase.  

\citet{Wetzel2013} used the fraction of quenched galaxies in local clusters to estimate this delayed-then-rapid timescale, arriving at a value of $2-6$\,Gyr, a value also corroborated by \citet{Fossati_2017} based on modelling quenched fractions in 3DHST. 
Hence the presence of mature clusters at $z>1.5$, less than 4.5\,Gyr after the Big Bang, implies that this quenching timescale is likely to evolve. Assuming that infalling pre-cluster galaxies and field galaxies have similar quiescent fractions, \citet{Balogh_2016} derived quenching timescales of 1.7\,Gyr for $z=1$ clusters, and \citet{Foltz_2018} extends this to $z=1.6$ to derive even shorter timescales of 1.1\,Gyr. In all of these studies the quenching timescale has a mass dependency, with the most massive galaxies quenching quicker than the lower mass galaxies.


The high quenched fractions in $z\simeq1$ clusters \citep{Newman_2014, Cook_2016, Vanderburg2020}, with their implied short quenching timescales, have important consequences for the predicted stellar ages of the quiescent cluster galaxies. If the high quenched galaxy fraction is caused primarily by assembly bias -- where galaxy formation started earlier in the protocluster compared to the field -- then the stellar ages of the quiescent cluster galaxies at $z\simeq1$ are predicted to be approximately a Gyr older than coeval field galaxies \citep{Vanderburg2020, Webb_2020}. On the other hand, if the high quiescent fraction was caused by rapid environmental quenching of infalling star-forming galaxies then the average stellar age of the quenched cluster galaxies is predicted to be 1.5 Gyrs younger than coeval quiescent field galaxies \citep{Webb_2020}, which quench over longer timescales.

It is the age dating studies of \cite{Gobat2008, Rettura2010, Lee-Brown2017, Webb_2020} that causes a conundrum for mature high-redshift clusters, since these studies find z $>$1 cluster galaxies are only marginally older than field quiescent galaxies. The stellar age difference is not large enough to account for the excess of quenched galaxies through assembly bias, but environmental quenching cannot be the dominant quenching mechanism since the derived short environmental quenching timescales are in direct contradiction to the older stellar ages of the cluster quiescent galaxies. Cluster-to-cluster variation cannot reconcile this problem as in some cases these contradicting results have been derived using the same clusters \citep[c.f.][]{Vanderburg2020, Webb_2020}.

In this study we address this apparent contradiction by exploring one of the underlying assumptions made when deriving the environmental satellite quenching timescales: that galaxies which fall into clusters have similar properties to \lq field\rq\ galaxies. Quenching timescales are usually calculated from the fraction of excess quenched galaxies in clusters with respect to a control field, i.e.~galaxies outside clusters or galaxies in the lowest density region of a survey. There are two reasons why this assumption may be unsound:

\noindent(i) The infalling galaxies may be "pre-processed" by other environmental influences before they fall into the cluster. The standard cosmological paradigm predicts that galaxy clusters form hierarchically. Small groups form first in the early Universe, which then merge into progressively larger systems.  So, before galaxies become satellites of a cluster they may travel through the modestly dense environments of groups and filaments \citep{DeLucia2012}, which may quench or alter the properties of the galaxies. This type of pre-processing is most commonly thought to act on galaxies which are satellites of another halo before they fall into the cluster. The prevalence of pre-processing at $z<0.8$ has been established by numerous observations of quenching and morphological galaxy transformations occurring out to several virial radii of massive clusters \citep[e.g.][]{Lewis2002, Gomez2003, Patel2011, Oemler2013, Haines2015, Bianconi2018, Just2019}, and probing the large scale structures around clusters for evidence of the effect of pre-processing and quenching has a long history \citep[e.g.][]{Kodama2004,Tanaka2006,Koyama2011}. There is also tentative evidence that pre-processing occurs even at $z\sim1.5$ \citep{Nantais2016}.

\noindent(ii) The progenitors of cluster galaxies form in protoclusters: an environment that was several times denser than the mean density of the Universe \citep{Chiang2013}. This  protocluster environment may alter the distributions of several properties of its member galaxies. Simulations predict that protocluster galaxies have a top heavy galaxy stellar mass function, started forming stars earlier, and are hosted by haloes that also have a top-heavy mass distribution \citep{Chiang2017,Muldrew_2018}. There is some observational evidence from $z>2$ protoclusters to back up these predictions \citep{Cooke2014, Willis2020}, and high and intermediatery density environments at high-redshift appear to accelerate galaxy growth \citep{Sobral2011, Hatch2011}. Therefore, a higher fraction of protocluster galaxies may be quenched compared to field galaxies, even if they are centrals of their own haloes.

The aim of this work is to determine whether the galaxies in the infall region of $0.8<z<1.4$ clusters have similar masses and quenched fractions as field galaxies. If they differ, we will examine how this difference affects the determination of the environmental quenching efficiency, and whether pre-processing or biased galaxy formation in the infall region can resolve the discrepancy between the old stellar ages and the apparent rapid quenching times of massive galaxies in high-redshift clusters.  

In Section 2 we introduce the $0.8<z<1.4$ clusters we use in this work and describe how we select infall, cluster and control galaxies from the data. In Section 3, we compare infall galaxies to cluster galaxies and control galaxies at the same redshift. 
We furthermore investigate the halo environment of the infall and control region by measuring the distribution of satellite galaxies around massive control and infall galaxies. We discuss our findings in Section 4 and  outline our conclusions in Section 5. 
We use AB magnitudes throughout and a $\Lambda$CDM flat cosmology with $\Omega_M=0.3$, $\Omega_\Lambda=0.70$ and $H_0=70$\,km\,s$^{-1}$\, Mpc$^{-1}$.

\section{Data and samples}

\begin{table}
\begin{center}
\begin{tabular}{||c c c c c c||}

\hline
Name   & RA   & Dec   &  Redshift  & $\sigma$   & \rv\   \\ [0.5ex] 
 & (J2000)     & (J2000)   &    & (km/s)  & (Mpc)  \\ [0.5ex] 
 \hline

SpARCS0034 & 8.675   & -43.132 & 0.867 & 700  & 0.58 \\ [2pt]
SpARCS0036 & 9.188   & -44.181 & 0.869 & 750  & 1.06 \\ [2pt] 
SpARCS1613 & 243.311 & 56.825  & 0.871 & 1350 & 1.54\\ [2pt]
SpARCS1047 & 161.889 & 57.687  & 0.956 & 660  & 0.91 \\ [2pt]
SpARCS0215 & 33.850  & -3.726  & 1.004 & 640  & 0.88 \\ [2pt]
SpARCS1051 & 162.797 & 58.301  & 1.035 & 689  & 0.84\\ [2pt]
SPT0546    & 86.640  & -53.761 & 1.067 & 977  & 1.15 \\[2pt]
SPT2106    & 316.519 & -58.741 & 1.131 & 1055 & 1.21 \\[2pt]
SpARCS1616 & 244.172 & 55.753  & 1.156 & 782  & 0.92 \\ [2pt]
SpARCS1634 & 248.654 & 40.364  & 1.177 & 715  & 0.85 \\ [2pt]
SpARCS1638 & 249.715 & 40.645  & 1.196 & 564  & 0.73 \\ [2pt]
SPT0205    & 31.451  & -58.480 & 1.323 & 678  & 0.85 \\[2pt]
SpARCS0219 & 34.931  & -5.525  & 1.325 & 810  & 0.9* \\ [2pt]
SpARCS0035 & 8.957   & -43.207 & 1.335 & 840  & 0.90 \\ [2pt]
SpARCS0335 & 53.765  & -29.482 & 1.368 & 542  & 0.69 \\ [2pt]

\hline
\end{tabular}
\caption{\label{tab:table-name}The 15 clusters from the GOGREEN and GCLASS samples that are used in this work. Column 4 provides the redshift of the cluster. Columns 5 and 6 provide the intrinsic velocity dispersion and radius in proper Mpc from \citet{Biviano_2021}, except for the cluster marked with * where we estimated the radius from $\sigma$ provided by GOGREEN DR1.}
\label{clusters}
\end{center}
\end{table}

\subsection{The GOGREEN and GCLASS cluster surveys}

We use data from the first public data release (DR1) of the Gemini Observations of Galaxies in Rich Early ENvironments (GOGREEN) and Gemini CLuster Astrophysics Spectroscopic Survey (GCLASS)
surveys\footnote{http://gogreensurvey.ca/data-releases/data-packages/gogreen-and-gclass-first-data-release} \citep{Muzzin2012gclass,Balogh_2017,Balogh_2021}, which contains photometric and spectroscopic data for 26 clusters and groups with redshifts between 0.85 and 1.50, and masses of at least $M_{200}\sim10^{13}M_{\odot}$. Three of these clusters were discovered using the Sunyaev-Zeldovich effect \citep{Sunyaev_Zeldovich1970} with the South Pole Telescope (SPT) \citep {Foley_2011, Stalder_2013,Sifon_2016}, whilst 14 were discovered using the red-sequence galaxy selection method as part of the Spitzer Adaptation of the Red-Sequence Cluster Survey (SpARCS; \citealt{Wilson2009, Wilson20092, Muzzin2009}), and nine groups in the COSMOS field were selected based on diffuse X-ray emission implying a well established intragroup medium \citep{Finoguenov_2010,Finoguenov_2007,George_2011}.

In this work, we only use the most massive clusters in the sample with intrinsic velocity dispersions $\sigma>500$km/s, which have dynamical masses of $>10^{14}$M$_\odot$. We make this selection because the size of the infall region of high-redshift clusters depends on the $z=0$ mass of the cluster \citep{Muldrew2015}, which correlates with cluster mass at the observed redshift. $\Lambda$CDM predicts that only 30\% of today’s cluster galaxies resided in the central cluster at z$\sim$1; the vast majority of galaxies lived around the cluster in a region that we refer to as the infall region \citep{Muldrew2015}. The most massive clusters today, of $M_{z=0}>10^{15}$\,M$_\odot$, had infall volumes that stretched out over a radial distance of $3-4$\,Mpc at $z=1$. The infall radii of more typical $M_{z=0}\sim10^{14}$\,M$_\odot$ clusters only extended $1.5-2.5$\,Mpc at $z=1$ \citep{Muldrew2015}. Since we wish to select the galaxy sample that will accrete onto the cluster by the present day, we select only the most massive $z=1$ clusters that are expected to have large infall surroundings.  
We list the 15 clusters used in this work in Table \ref{clusters}.

The GOGREEN photometric catalogues contain deep photometry from $u$ to $4.5\mu$m wavelengths. Descriptions of the photometric data and the image processing are described in \citet{Vanderburg_2013}, \citet{Vanderburg2020} and \citet{Balogh_2021}. Accurate relative colour measurements were obtained for each source by \citet{Vanderburg2020} using PSF-homogenised image stacks. These colours are used to identify stars using $uJK$ colour criteria and we use the \lq star\rq\ classification included in the data release and remove all sources classified as \texttt{star = 1}.

The DR1 photometric redshifts and rest-frame $U-V$, $V-J$ colours were estimated by \cite{Vanderburg2020} using the \texttt{EAZY} code (Version May 2015; \cite{Brammer_2008}), by fitting the photometry to spectral energy distribution (SED) templates from the PEGASE model library \citep{Fioc1997} with an additional red galaxy template from \cite{Maraston2005}. The photometric redshift uncertainty for galaxies with stellar masses greater than $10^{10}$M$_\odot$ is  $0.048(1+z)$ with 4.1\% outliers. 

Galaxy stellar masses provided by DR1 were obtained by fitting the photometry with stellar population synthesis models from \citet{Bruzual2003} using the FAST code \citep{Kriek2018}, assuming solar metallicity, the \cite{Chabrier2003} initial mass function and using the \cite{Calzetti2000} dust law. The star-formation history of these models were limited to exponentially declining functions, which are known to underestimate the stellar mass by up to 0.3\,dex compared to non-paramaterised models \citep{Leja_2019}. Throughout this work we limit our analysis to galaxies with stellar masses $>10^{10}M_{\odot}$, which is the 80\% completeness limit as determined by \cite{Vanderburg2020}. Following \cite{Williams2009} we used the rest-frame $U-V$, $V-J$ colours to classify galaxies as star-forming or quiescent. 
 We used the following criteria:
\begin{equation}
  U-V > 1.3 \And  V-J < 1.5 \And  U-V > 0.88 (V-J) + 0.59, 
  \label{eqn:colour}
\end{equation}
as defined by \citet{Muzzin2013}.

The cluster centre is defined as the position of the brightest cluster galaxy (BCG) within each cluster, where the BCG is the most massive galaxy within 500\,kpc of the main galaxy overdensity which has a redshift that is consistent with the cluster. 
We used the velocity dispersions, $\sigma$, and $R_{200}$ calculated by \citet{Biviano_2021}, except for the one cluster not in their sample, SpARCS0219, where we estimated the radius through $R_{200}=\sqrt{3}{\sigma/10 H(z)}$, where $H(z)$ is the Hubble parameter at the cluster's redshift \citep{Schneider2006}.

\subsection{Classifying cluster, infall and control field galaxies}
\begin{figure*}
\centering
\includegraphics[width=2\columnwidth]{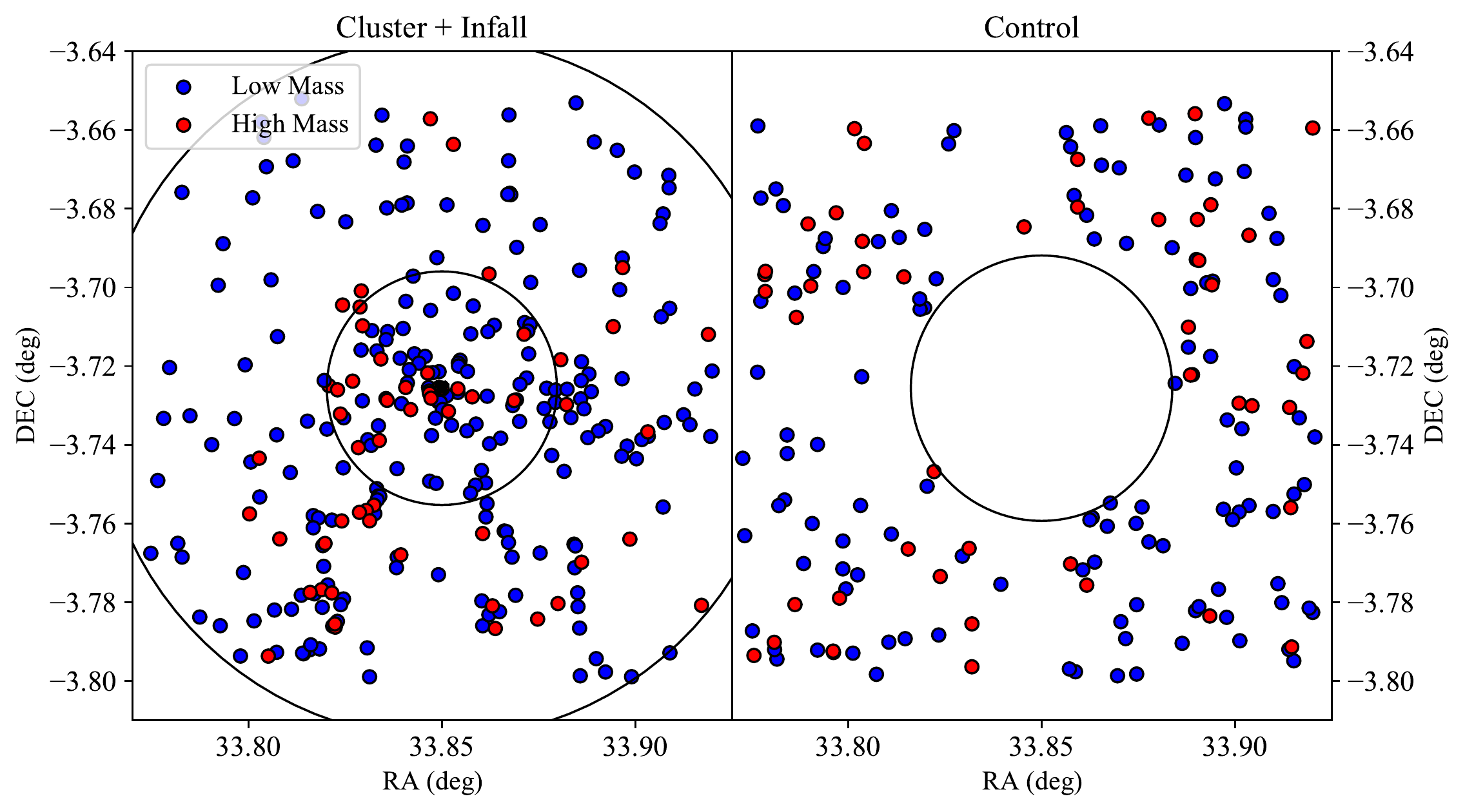}
\caption[]{Positions of galaxies in our cluster and infall samples (left panel) and control sample (right panel) within the SpARCS215 field. Blue points mark the positions of lower mass galaxies with  $\mathrm{9.75<log(M_{*}/M_{\odot})<10.8}$. Red points locate galaxies with $\mathrm{log(M_{*}/M_{\odot})>10.8}$. In the left panel the inner circle marks the \rv\ boundary. Galaxies within the circle (and $|\Delta z| /(1+z)<0.08$) are classified as cluster members. The outer circle marks 3\rv: infall galaxies lie between the inner and outer circle (and $|\Delta z| /(1+z)<0.08$).  The circle in the right panel marks 1\,Mpc from the BCG. The control galaxies are distributed outside this circle with redshifts in the range $0.15<|\Delta z| /(1+z)<0.3$.}
\label{RA_DEC}
\end{figure*}

We define three samples of galaxies: cluster members, infall members, and a control sample containing galaxies in neither of these environments.
We make a distinction between the cluster galaxies that are orbiting the potential and infall galaxies that are gravitationally bound, but not yet in stable orbits, but rather still lie on dominantly radially infalling paths. According to simulations, galaxies within a projected radius of approximately $R_{200}$ are predominantly on orbital paths, whilst those at greater projected radii are predominantly on their first infall towards the cluster \citep{Haines2015}. 
However, a complication arises due to the presence of backsplash galaxies. 

Backsplash galaxies refer to the galaxy population that have passed through the central region of the cluster and are about to turn around to make their second pass of the core \citep{DiemerKravtsov_2014}. Although backsplash galaxies make up over half of the galaxies found between $R_{200}$ and $2R_{200}$ around $z=0$ massive clusters, only 10\% of the galaxies between $R_{200}$ and $2R_{200}$ of their progenitors at  $z=1$ are backsplash \citep{Haggar_2020}. Thus $R_{200}$ is an appropriate divide that relatively cleanly separates cluster galaxies from infall galaxies for massive clusters at $z=1$. 

\begin{figure*}
\centering
\includegraphics[width=2\columnwidth]{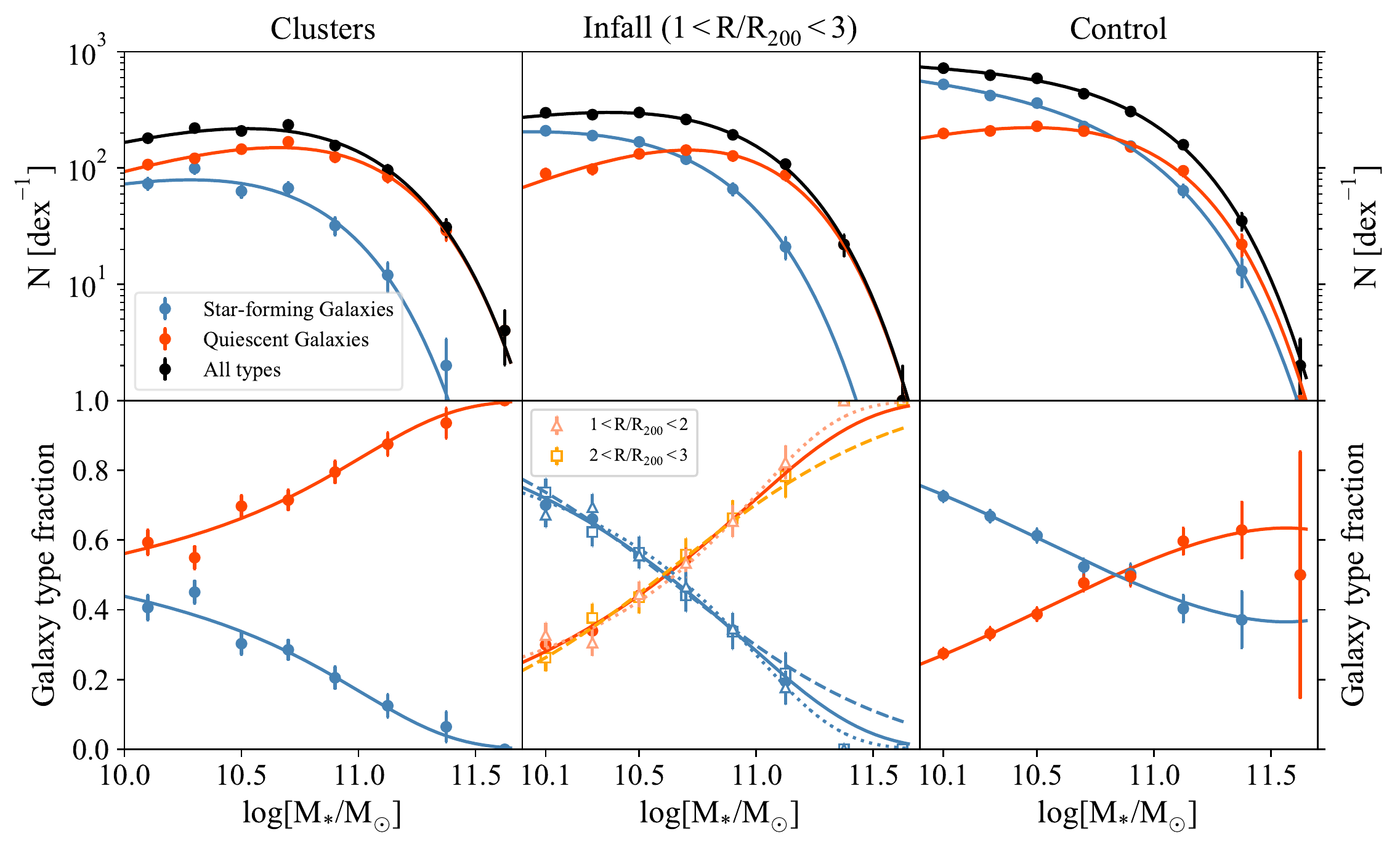}
\caption[]{Galaxy stellar mass functions (top row) and (bottom) fractions of star-forming (blue) and quiescent (red) galaxies within the cluster sample (left panels), infall (middle panels) and control sample (right panels). Errors on the black points (for all galaxy types) are Poissonian, and binomial errors are used on the subclasses in red and blue. The solid lines are the most likely Schechter functions that fit the data. The dotted (dashed) line in the bottom middle panel displays the fraction of each galaxy type in the inner $1<R/$\rv$<2$ (outer $2<R/$\rv$<3$) regions of the infall. There is no significant difference in the quiescent fractions between these regions so backsplash galaxies do not impact these results.}
\label{SMF}
\end{figure*}

We further limit the infall population to those galaxies within $3R_{200}$, which in most cases correspond to $3-4$\,Mpc (physical). This distance corresponds to the maximal radial extent of 90\% of the galaxies that will fall into the cluster by $z=0$ \citep{Muldrew2015}. We note that not all GOGREEN or GCLASS fields extend as far out as $3R_{200}$. This does not affect our results since will show (in Fig.\,\ref{SMF}) that the infall population looks similar in the two radial bins of $1<R/R_{200}<2$ and $2<R/R_{200}<3.$

 We use photometric redshifts and projected radial distances to define our samples. We define cluster members as galaxies within a projected virial radius (hereafter, \rv) of the cluster centre and photometric redshifts within $0.08(1+z)$ of the cluster's mean redshift, i.e. $|z_{\rm phot}- z_{\rm cl}|/(1 + z_{\rm phot})<0.08$. We define infall members as galaxies that lie between \rv\ and 3\rv\ from the cluster centre and have $|z_{\rm phot}- z_{\rm cl}|/(1 + z_{\rm phot})<0.08$.  The |$\Delta z| = 0.08(1+z)$ interval was chosen to allow us to directly compare our results with \vdBt.

The |$\Delta z$| window of $0.08(1+z)$ encompasses 39 cMpc at $z=1$ so the cluster and infall regions only reside within a fraction of the volume selected. The galaxy overdensity is expected to be higher in the cluster than in the infall region, thus we expect a higher fraction of interloper contamination in the infall sample. Furthermore, the photometric redshift uncertainties lead to cluster and infall galaxies being scattered out of the cluster and infall samples. This can be corrected if a large and representative sample of the galaxies have a spectroscopic redshift \citep[cf.][]{Vanderburg2020}. The cluster sample have sufficient spectra to perform this correction, but the infall sample does not. Therefore, we do not apply corrections to either the cluster or the infall sample. We note that the typical size of the corrections performed by \vdBt\ is of order of 20\%, or 0.1 dex in the log-log plot of a stellar mass function. Therefore these corrections are relatively minor and are not expected to  significantly affect our overall results. Nevertheless, we test the robustness of our results subject to variations in the |$\Delta z$| membership selection in Appendix \ref{appendix:a}.

In addition to the cluster and infall galaxies we also create a \lq control\rq\ galaxy sample consisting of galaxies within the GOGREEN and GCLASS fields that lie close to the redshift of the cluster, but not within the cluster or infall volume. Galaxies in this control sample have not experienced the same environmental influences as the cluster and infall galaxies, but are subject to similar observational selection biases.

We define the control sample as those galaxies with photometric redshifts in the redshift interval $0.15<$|$z_{\rm phot}- z_{\rm cl}$|$/(1 + z_{\rm phot})<0.3$, and lie at least 1\,Mpc away from the cluster centre, where $z_{\rm cl}$ is the redshift of the cluster in each field. The inner |$\Delta z$| limit was chosen such that galaxies were at least three times the photometric redshift uncertainty away from the cluster's redshift, whilst the outer boundary was designed to limit the control galaxy sample to a similar redshift range as the cluster and infall samples. The spatial distribution of cluster, infall and  control  galaxies are displayed for an example cluster, SpARCS0215, in Figure \ref{RA_DEC}. In total, the cluster galaxy sample contains 1113 members, the infall galaxy sample contains 1442 members, whilst the control sample contains 2632 members to a mass limit of $10^{10}$\Msun. 

The importance of selecting a control sample with the same data as the environmentally processed galaxy sample has been emphasised by \citet{Papovich_2018}, however our resulting control sample is relatively small and therefore subject to Poisson noise. Furthermore, in the fields containing $z\sim1.3$ clusters our control sample includes galaxies up to $z\sim2$ due to the necessity of selecting control galaxies at a significantly different redshift range from the infall region. This means the completeness corrections we apply to the low mass control field galaxies (see next section) are larger than for the infall and cluster galaxies. This may influence our result so we test the robustness of our results subject to our control sample selection in Appendix \ref{appendix:b}.

\section{Results}
\subsection{Galaxy stellar mass functions}
\label{sec:GSF}

We begin our investigation of the galaxies in the infall regions surrounding $z\sim1$ clusters by examining the galaxy stellar mass function (SMF). 
We combine all the galaxies in the 15 fields, splitting by cluster, infall and control regions, then count the number of galaxies in each sample in mass bins of 0.2\,dex in the range $10<\log[M/M_{\odot}]<12$. We further divide the galaxies in each sample into star-forming and quiescent galaxy types and construct stellar mass functions for each type of galaxy.

We apply completeness correction factors for undetected sources in each mass bin. Each field has a slightly different detection limit, so we used the $\alpha =1$ completeness curves given in Appendix A.1 of \vdBt, shifted to the $Ks$ 80\% limiting magnitude appropriate to the data the galaxy is extracted from (see Tables 1 in \citet{Vanderburg_2013} and \vdBt, respectively). We assign each cluster, infall and control galaxy a weight of 1/completeness($Ks$) according to their $Ks$ magnitude. These completeness corrections are relatively minor: $\sim3$\% in the lowest mass bin for the infall and cluster samples, and $\sim16$\% for the lowest mass bin of the control sample.  We apply these weights in each mass bin to derive the corrected stellar mass functions, which are  presented in Figure \ref{SMF} with uncertainties assuming a Poisson distribution in each mass bin.

\begin{figure}
\includegraphics[width=1\columnwidth]{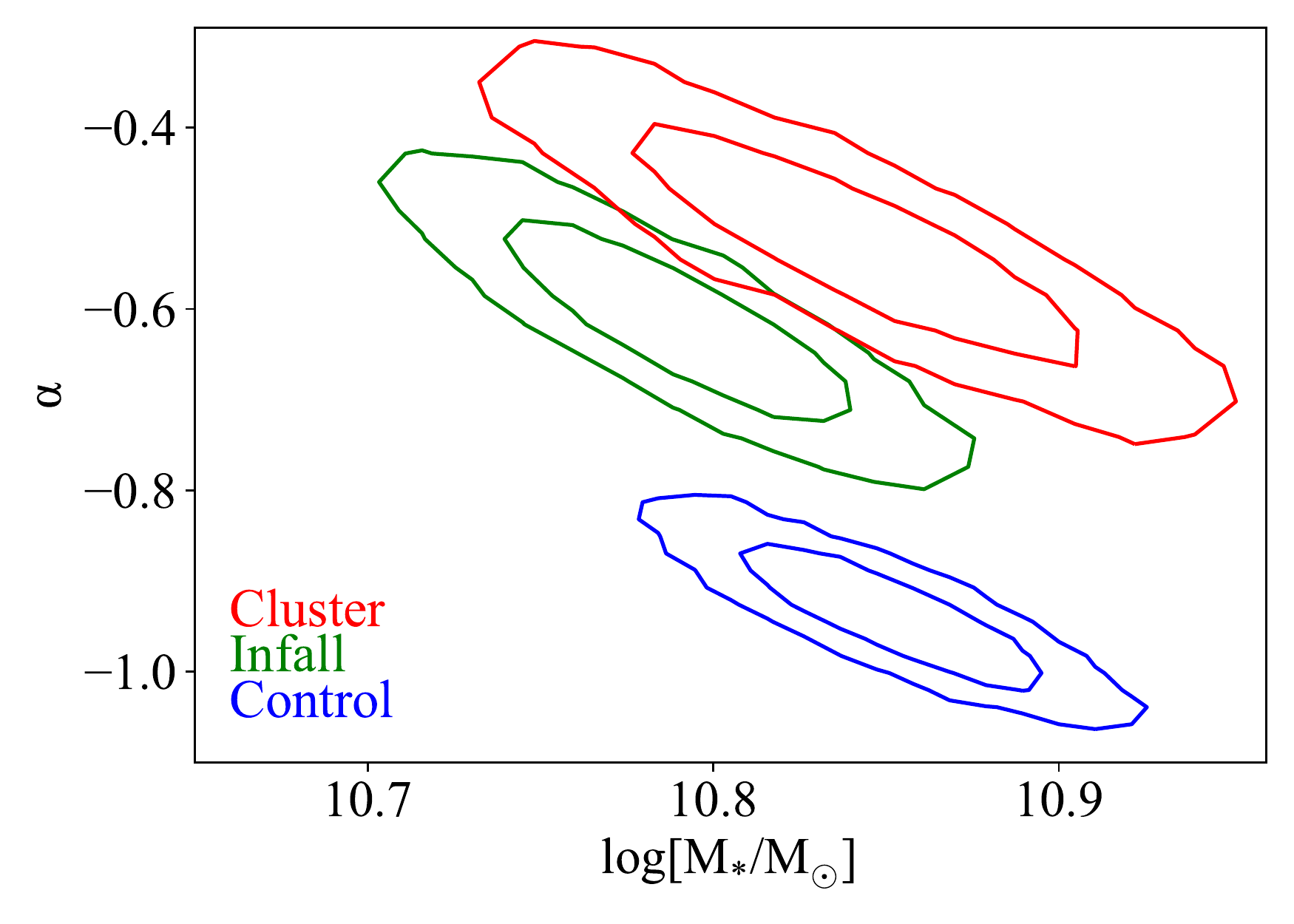}
\caption[]{The most likely characteristic mass, $M_{*}$, and low mass slope, $\alpha$ for the cluster (red), infall (green) and control (bue) samples. The contours mark 1 and 2$\sigma$ for each population. The low mass slope of the control sample differs by more than 2$\sigma$ meaning both the cluster and infall samples have top-heavy stellar mass functions.}
\label{SMF_parameters}
\end{figure}

We fit the SMFs with  Schechter functions \citep{Schechter1976}, using the python module \texttt{emcee} \citep{emcee2012} that uses the Markov chain Monte Carlo (MCMC) technique to find the characteristic mass (M$_{*}$), low-mass slope ($\alpha$) and normalisation that maximises the likelihood of fitting the data. We overlay the Schechter fits on the raw data in Figure \ref{SMF} showing that both the full population and the star-forming/quiescent subsets in all three samples are well fit by Schechter functions. The most likely values for $\alpha$ and M$_{*}$ for all the galaxies in the cluster, infall, and control samples are displayed in Figure \ref{SMF_parameters}, which shows that the shape of the Schechter functions for each sample differ by more than $2\sigma$. 

We are unable to measure the volumes of the cluster and infall regions so a comparison of the normalisations of the stellar mass functions does not reveal anything physically meaningful. We are therefore limited to comparing the shapes of the stellar mass functions. The characteristic mass in all 3 samples have similar values of M\,$\sim10^{10.83}$\Msun, but the low-mass slopes of the infall and cluster samples are shallower than the control sample. From Figure 2 we see that the cluster and infall samples contain higher ratios of massive to low mass galaxies than the control field. Hence the cluster and infall environments contain a relative excess of massive galaxies.

A comparison between the SMFs of our cluster and control sample agrees qualitatively with \vdBt's comparison of GOGREEN clusters and the UltraVISTA  field.  We obtain results that agree within 1$\sigma$ for the cluster sample, but the low mass slope of our control sample disagrees slightly with the UltraVISTA calculated in \vdBt\ due to our choice of fitting large mass bins (to mitigate  the uncertainty in stellar mass) rather than without binning the data (as is done in \vdBt). We find the low mass slope of UltraVISTA matches our control sample when it is calculated using the same large mass bins we use in this work. 

\subsection{Quiescent galaxy fraction}

\begin{figure}
\includegraphics[width=1\columnwidth]{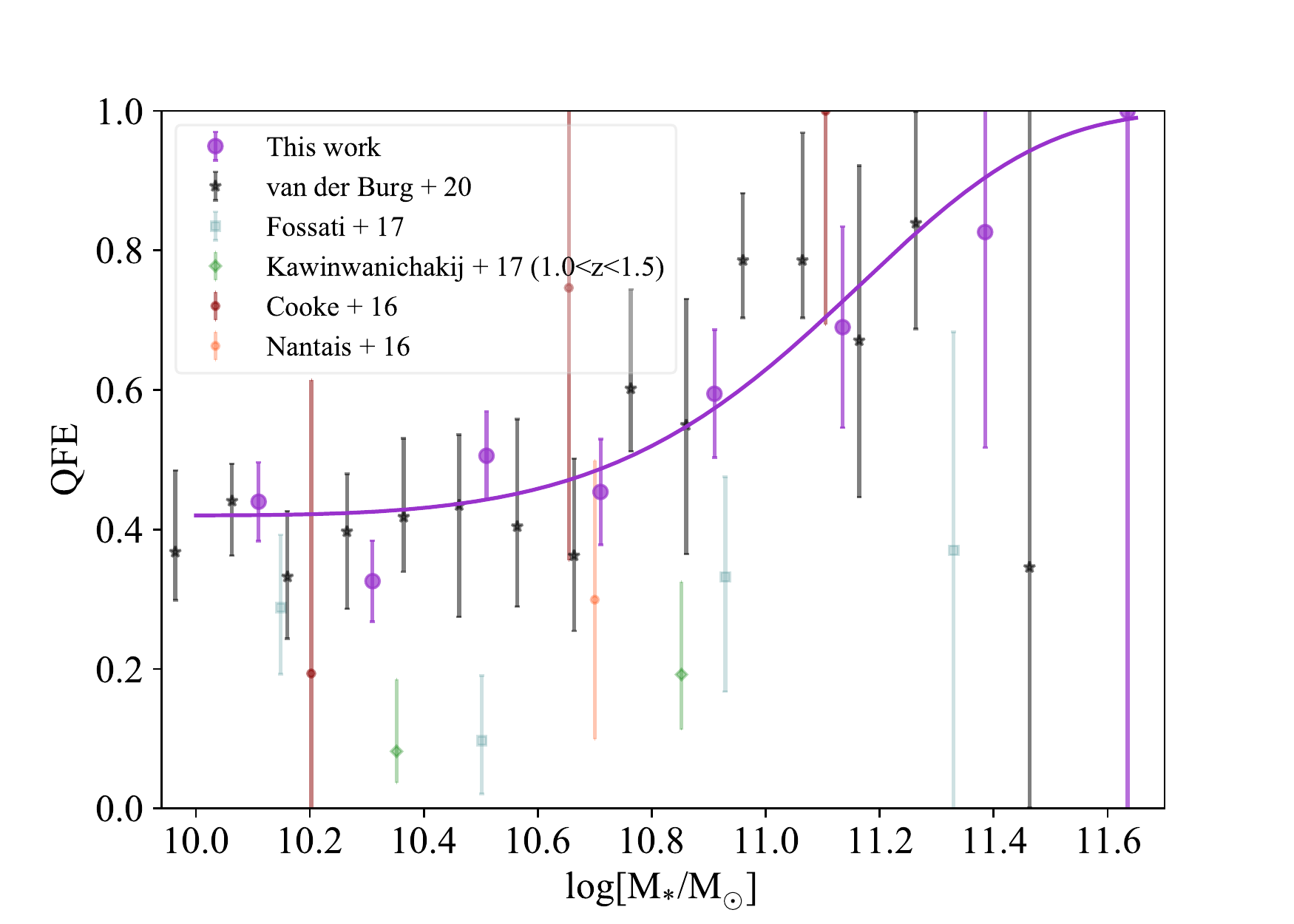}
\caption[parameters]{The excess of quiescent galaxies in the cluster compared to the control (purple points and solid line). We find a favourable comparison to literature results of a similar redshift and cluster mass: \cite{Vanderburg2020} (black), 
\cite{Nantais2016} (pink)
\cite{Fossati_2017} (blue), \cite{Kawinwanichakij_2017} (green), and \cite{Cook_2016} (dark pink) data. QFE = 0 occurs when there are no excess quenched galaxies in the cluster within the mass bin, whereas QFE = 1 occurs when all the star-forming galaxies in a mass bin are quenched in the cluster.}
\label{QFE_remco}
\end{figure}

We calculate the fraction of galaxies in each mass bin that are quiescent and show the results in the bottom panels of Figure \ref{SMF} with binomial uncertainties. 
To test whether the infall sample is contaminated by backsplash galaxies, we recalculate the quiescent fraction in two subregions: $1<$\,R/\rv$<2$ and $2<$\,R/\rv$<3$. As shown in bottom-middle panel of Figure \ref{SMF}, these quiescent fractions are consistent within uncertainties in all mass bins, even though the fraction of backsplash galaxies in the $2<$\rv\ $<3$ sample is predicted to be larger than in the $1<$\rv\ $<2$ sample. We argue, therefore, that backsplash galaxies are unlikely to significantly contaminate the infall sample.

All three galaxy samples present a clear trend of increasing quiescent fraction with increasing stellar mass, in agreement with studies at both low and high redshift \citep[e.g.][]{Peng2010}. 
The gradient of this trend in the cluster sample is similar to the control sample, but the normalisation is higher. 
However, the infall sample has a steeper gradient with stellar mass than either cluster or control sample. The most massive infall galaxies share a similar high quiescent fraction as the massive cluster galaxies, but the lowest mass infall galaxies share a similar low quiescent fraction as the control field galaxies. 
Therefore, the quenching of infall galaxies has a different dependency with stellar mass than either the cluster or control galaxies.

In order to examine the mass dependency of the quiescent fractions in more detail we calculate  the Quenched Fraction Excess (QFE) for each sample. The QFE is known in the recent literature by a variety of names, such as the conversion fraction or the quenching efficiency. First defined by \citet{vandenBosch_2008}, it calculates the excess fraction of quiescent galaxies in one sample relative to another. The samples can be galaxies within different mass bins, or within the same mass bin but in different environments (as in this work). We use the term QFE rather than conversion factor since  the environmental samples we compare may not be related to one another in a evolutionary sequence. For example, cluster galaxies do not evolve from the control field galaxies; the progenitors of the cluster galaxies are infall galaxies. 

\begin{figure}
\includegraphics[width=1\columnwidth]{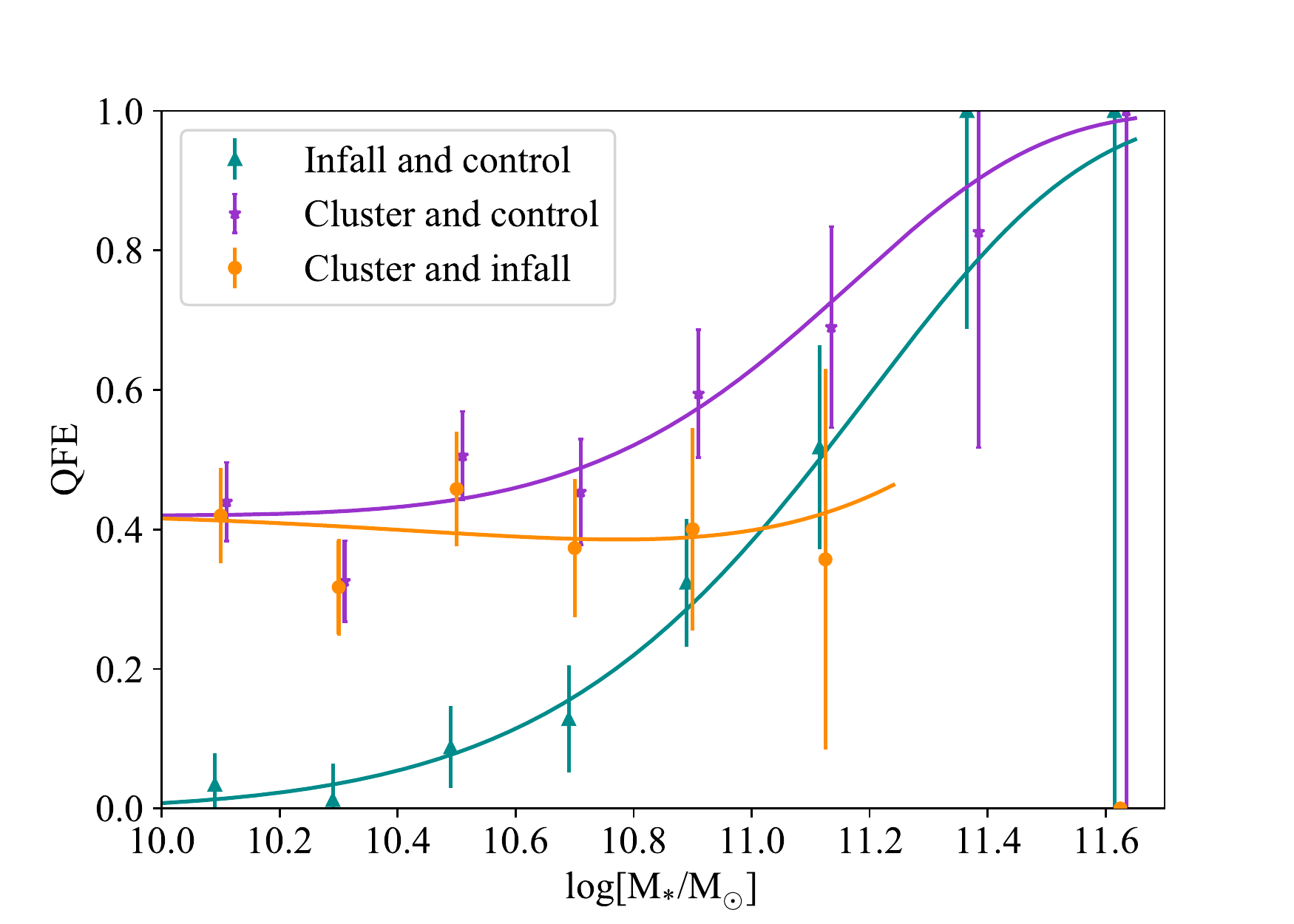}
\caption[parameters]{The orange data points show that the excess quiescent fraction in the cluster sample compared to the infall sample (QFE$_{cl-inf}$) is approximately constant across all stellar masses up to $10^{11.2}$\Msun. Above $10^{11.2}$\Msun\ there are no excess of quiescent galaxies in the cluster compared to the infall therefore above this limit QFE$_{cl-inf}=0$. The aquamarine data points show the excess quiescent fraction in the infall sample compared to the control sample (QFE$_{inf-con}$), which has a strong mass-dependency. The solid lines are obtained from the Schechter fits to the stellar mass functions rather than fits to the data points.}
\label{QFE}
\end{figure}

We define the QFE in each stellar mass bin through:  

\begin{equation}
    QFE_{2-1} = \frac{f_{{\rm q},2} - f_{{\rm q},1}}{1-f{_{\rm q,1}}},
\label{QFE_eq}
\end{equation} 
where $f_{{\rm q},1}$, $f_{{\rm q},2}$ is the quiescent fraction in a stellar mass bin in environment $1$ and $2$, respectively. A QFE of zero implies there is no excess of quenched galaxies in one environment compared to the other. A high QFE means that many of the star-forming galaxies observed in environment 1 must be quenched in environment 2. In the extreme case of QFE = 1, all of the galaxies that are star-forming in environment 1 would be quenched were they to reside in environment 2.

Figure \ref{QFE_remco} displays QFE$_{cl-con}$, which measures the excess quiescent fraction in the cluster sample compared to the control sample.  We show that the QFE$_{cl-con}$ we measure from our data is quantitatively similar to that of \vdBt, who use the COSMOS/UtraVISTA field as their control field. We take this as reassurance that although our control sample is smaller, and therefore more prone to Poisson noise, it is still sufficient to produce reliable results. 

We also compared our results for the QFE with other results in the literature. All studies agree that the QFE of massive galaxies is larger than the lower mass galaxies, but there is significant variation in the absolute values. We obtain good agreement with studies that calculate the QFE using high-redshift clusters \citep{Cook_2016, Nantais2016}, but studies that divide a large field into density bins  obtain  systematically lower QFEs \citep{Fossati_2017, Kawinwanichakij_2017}. It is likely that the cluster QFE studies are a more direct probe of satellite quenching whereas the field studies have greater contamination by isolated galaxies and  therefore the absolute values are not directly comparable.

Having shown our results for the cluster sample are consistent with the literature, we now turn to the main purpose of this work: the infall galaxies. We derive the QFE of the cluster sample compared to the infall sample (QFE$_{cl-inf}$) and show the results in Figure \ref{QFE}. In this case the QFE$_{cl-inf}$ can be considered a conversion factor. The infall galaxies are the progenitors of the cluster galaxies, so  QFE$_{cl-inf}$ is the fraction of infall galaxies that must be quenched when they fall into the cluster.  We also show the QFE of the infall sample compared to the control sample (QFE$_{inf-con}$) in Figure \ref{QFE}. In this case, the QFE should not be considered a conversion factor: control galaxies are not in an evolutionary sequence with the infall galaxies.  This is because we have selected the infall region such that it coincides with the region of the protocluster that will ultimately collapse to form the $z=0$ cluster. Hence galaxies within the infall regions were formed within the large-scale overdensity of protoclusters. The control galaxies lie outside this special overdense region and typically formed in a lower density large-scale environment. Furthermore, at $z\sim1$, many of the galaxies in the infall region have only recently reached the turn-around point of collapse in the protocluster and so they are just starting their infalling orbits onto the cluster core \citep{Muldrew2015}. Therefore, the infall regions abutting $z\sim1$ clusters are not fed by galaxies from the control environment at $z\sim1$ and QFE$_{inf-con}$ simply displays the difference in the quenched fraction between these galaxy samples.

Figure \ref{QFE} shows that the cluster contains a relatively constant excess of quenched galaxies across the entire mass range probed (M\,$>10^{10}$\Msun) compared to the infalling sample. This implies that the process that quenches the infall galaxies when they fall into the cluster does not have a strong mass-dependency. By contrast, we observe a strong gradient in the QFE$_{inf-con}$ with stellar mass. There is a large excess of massive quenched galaxies in the infall regions compared to the control. Whereas there is almost no excess of quenched galaxies with masses $<10^{10.5}$\Msun.

We have already shown that any backsplash galaxies present do not affect the quiescent fraction of the infall sample (see middle panel of Figure \ref{SMF}), so they do not affect the QFEs displayed here. In Appendix \ref{appendix:a} we show that the choice of |$\Delta z$| for selecting infall and cluster galaxies  does not affect the QFEs presented here, and in Appendix \ref{appendix:b} we show that the QFE are not sensitive to the selection criteria of the control sample. 

The QFEs suggest that massive infall galaxies are quenched more efficiently than similar mass galaxies in the control sample. Since the stellar masses of these galaxies are similar, we hypothesise that the environments are different and a process that depends on environment is responsible for this enhanced quenching rate. In the following section we compare the environments of the massive infall and control galaxies.
\begin{figure*}
\includegraphics[width=2\columnwidth]{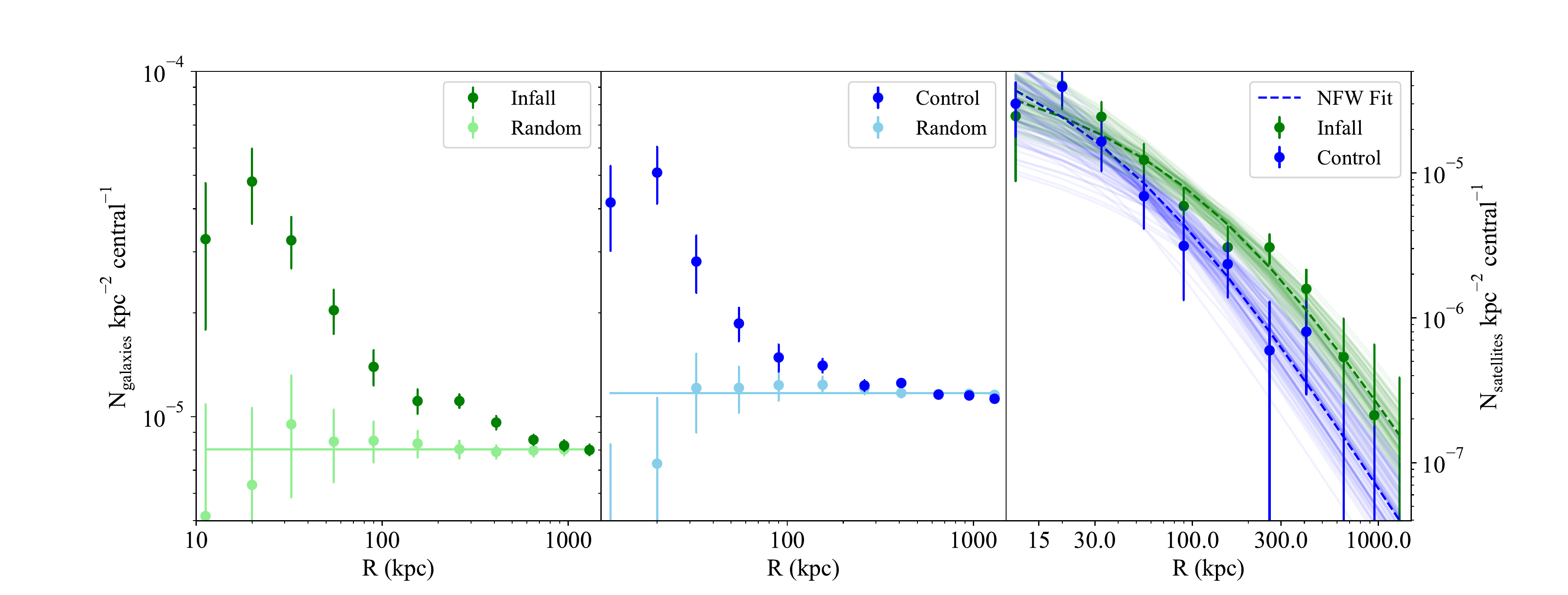}
\caption[]{Number of (satellite) galaxies per area per central galaxy as a function of projected distance. The left panel displays, in dark green, the  density of infall galaxies located in projected radial bins around $>10^{10.8}$\Msun\ galaxies in the infall sample. In light green we display the galaxy density in radial bins around random position in the infall region to measure the expected level of contamination; the solid line is the median. The middle panel displays a similar analysis as the left panel, but using control galaxies of $>10^{10.8}$\Msun. Subtracting the contamination (straight solid lines) from the dark blue and green points results in the excess satellite galaxy density, which is shown in the right panel. The dashed lines are the most likely NFW fit to the satellite galaxy distributions, whilst the transparent lines are 2\% of the samples from the MCMC chain selected by random.}
\label{satellites}
\end{figure*}

\subsection{Halo properties of massive infall galaxies}

In $\Lambda$CDM cosmogonies, massive dark matter halos grow by the assimilation of smaller halos. Galaxies that reside in the smaller haloes at the time of accretion become satellite galaxies in the massive halo. Several studies have shown that these satellite galaxies settle into a distribution around the central galaxy according to an NFW \citep{NFW_1996} profile \citep{Nierenberg_2011, Nierenberg_2012, Tal_2012,Tal_2013, Wang_2012, Watson_2012, Wang_2014,Kawinwanichakij_2014}. 
Hence the number and distribution of satellites around massive infall and control galaxies provides a means to estimate the size and mass of their dark matter haloes.

For each $M>10^{10.8}$\Msun\ galaxy in the control and infall samples we measure the number of lower-mass neighbouring galaxies (to a mass limit of $10^{10}$\Msun) within projected radial bins from 7.5\,kpc to 1.5\,Mpc. We exclude the area of $R<1$\,\rv\ and $R >3$\,\rv\ for infall galaxies and $R<1$\,Mpc for control galaxies, and account for bright stars in each field using the mask images. 
We apply completeness correction factors for undetected sources in each radial bin since each field has a slightly different detection limit. We assign each infall and control galaxy a weight of 1/completeness($Ks$) according to their $Ks$ magnitude, as described in Section\,\ref{sec:GSF}, then apply these weights in each radial bin to derive the corrected radial distributions. We calculate uncertainties for the galaxy density in each radial bin by repeating the calculation 100 times using bootstrap with replacement of the galaxy samples and taking the standard deviation of each radial bin.

We also calculate the radial distribution of galaxies around a similar number of random points in each field. For the infalling population we ensure the random positions are distributed at similar distances from the cluster centre as the massive infalling galaxies. We assign each random point a mass M\,$>10^{10.8}$\Msun\ from the stellar mass distribution. We then repeated the process 100 times to obtain the mean galaxy density and standard deviation in each radial bin surrounding these random positions. This provides the level of contamination due to non-associated galaxies that can be expected within each radial bin.

The density of galaxies surrounding the massive infall and control galaxies are shown in the left and central panels of Figure \ref{satellites}, respectively. We also show the density of infall and control galaxies around random positions, which is consistent with a constant across all radial bins. There is a strong excess of galaxies surrounding both massive infall and control galaxies out to 1\,Mpc and 500\,kpc, respectively. 

We obtain the satellite galaxy density by subtracting the median density of galaxies around random positions from the number of neighbours around massive infall and control galaxies. We display the density of satellite galaxies in the right panel of Figure \ref{satellites}, which shows that massive infall galaxies host significantly more satellite galaxies than similar mass control galaxies in almost all radial bins. To quantify the excess of satellite galaxies we measure the total galaxy excess within $7.5<R<500$ kpc for each sample. We detect on average $1.3\pm0.1$ satellites per massive infall galaxy, whereas there are only $0.6\pm0.1$ satellites per massive control galaxy. Thus we find there are twice as many satellites around the infall galaxies than around the control galaxies (significant at a $4.7\sigma$ level). 

To quantify how this difference in satellite density translates into a difference in dark matter halo mass we fit the radial distributions with projected NFW profiles. We use the  projected profiles from \citet{Bartelmann_1996}:

\begin{equation}
    \Sigma (x) = 
        \begin{cases}
        n(x^{2}-1)^{-1} \left( 1- \frac{2}{\sqrt{x^{2}-1}} arctan \sqrt{\frac{x-1}{x+1}} \right) & (x>1) \\
        
        n/3 & (x=1) \\
        
        n(x^{2}-1)^{-1} \left( 1- \frac{2}{\sqrt{1-x^{2}}} arctanh \sqrt{\frac{1-x}{1+x}} \right) & (x<1) \\       
        
        \end{cases}
\end{equation}

where $x=r/r_s$, $r_s$ is the NFW scale radius and $n$ is the normalization. We do not include the inner most data point in the fit as it is clear from Figure \ref{satellites} that the density is reduced and therefore does not conform to an NFW fit. This may be due to the difficulty of identifying galaxies so close to the central or due to the effects of dynamical friction and satellite cannibalism. 

We use the MCMC technique with the python module \texttt{emcee} \citep{emcee2012} to find the $r_s$ and normalisation that have the maximum likelihood to fit the observed profiles. We overlay the most likely NFW fits on the satellite distributions in the right-hand panel of Figure \ref{satellites}, which shows that both control and infall satellites conform to NFW profiles at projected radii $>30$\,kpc. The profile within $30$\,kpc\ is generally better fit by a power-law rather than the NFW-profile \citep[e.g.][]{Tal_2012,Kawinwanichakij_2014}. 

\begin{figure}
\includegraphics[width=1\columnwidth]{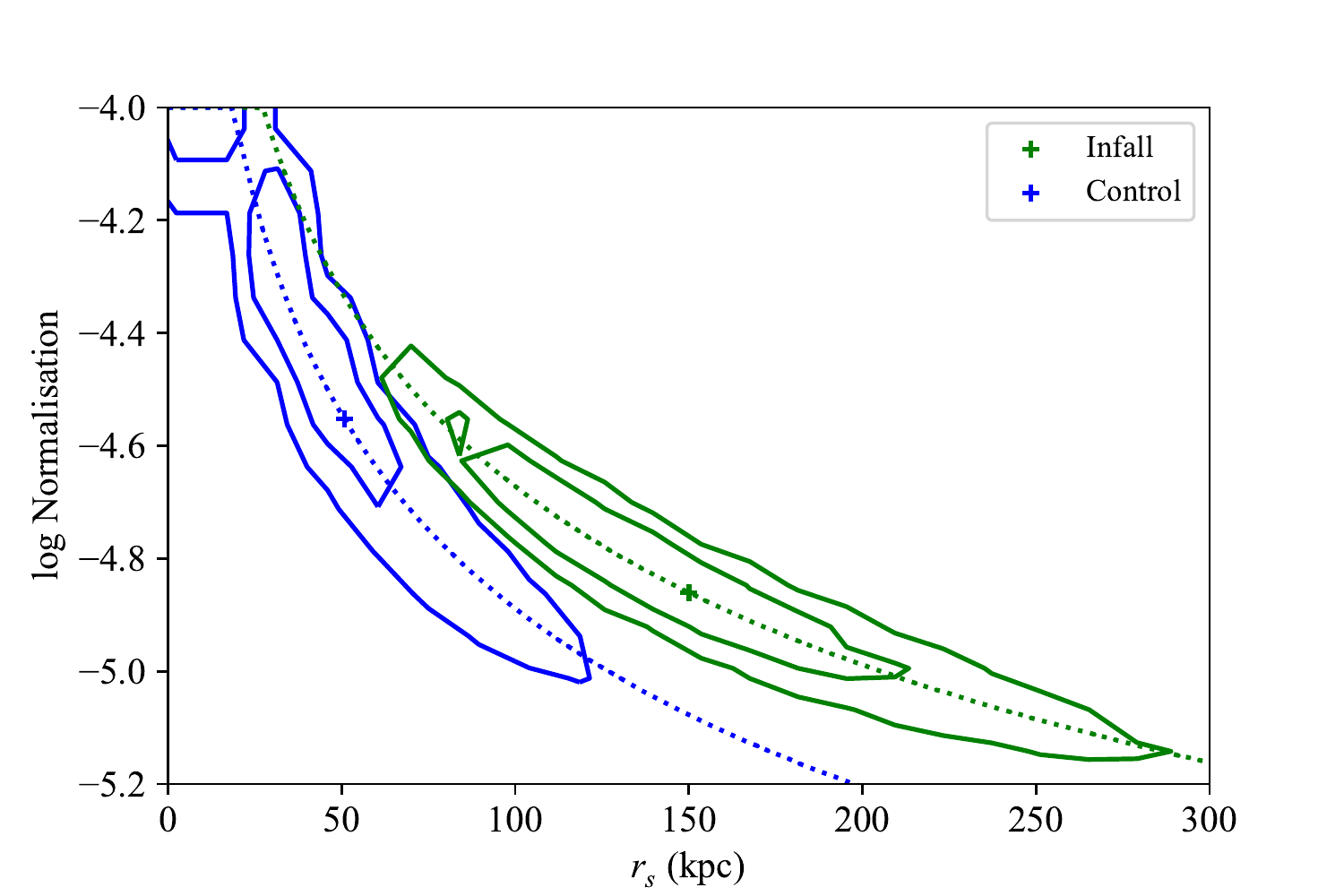}
\caption[]{Distribution for the scale radius, $r_s$, and normalisation, $n$ obtained from an  MCMC analysis of the projected satellite galaxy distributions for the control sample (blue) and for the infall (green). Contours mark 50\% and 80\% significance and the dotted lines mark the region of constant integrated galaxy density consistent with the number of observed satellites in each sample.}
\label{satellites_parameters}
\end{figure}

The 50 and 80\% significance contours for $r_s$ and the normalisation are displayed in Figure \ref{satellites_parameters}.  The optimal NFW parameters for the infall sample differ from the control sample at 80\% significance. Marginalising over the normalisation parameter, we find the scale radius for the haloes surrounding the massive infall galaxies is larger than the haloes hosting the control galaxies: $r_s=150_{-56}^{+75}$\,kpc  for the infall galaxies and $r_s=50_{-22}^{+52}$\,kpc for the control galaxies ($1\sigma$ uncertainties). 

We may lose some satellites of massive galaxies that have photometric redshifts close to the edges of our redshift intervals. We therefore repeated the above tests, but  using infall and control galaxies within redshift intervals that are $0.05(1+z)$ wider than the massive galaxy redshift intervals in each sample. We find similar results: there are on average $1.6\pm0.1$ satellites per massive infall galaxy, whereas there are only $0.9\pm0.1$ satellites per massive control galaxy. Both sets of satellite galaxies are distributed in NFW-like profiles that are consistent with the scale radii presented above.

Both the number of satellites (or richness) within a halo and the scale radius of the halo are proxies for halo mass  \citep{NFW_1996,Papovich2016}, therefore our results suggests that the massive infall galaxies typically occupy higher mass haloes than control galaxies of the same stellar mass.

\section{Discussion}

An important finding of our study is that there are differences between the infall and control galaxies. The infall sample has a relative excess of massive galaxies  (Figures \ref{SMF} and \ref{SMF_parameters}) and its quiescent fraction has a steeper gradient with stellar mass than the control sample. 
This results in a higher fraction of massive infall galaxies being classified as quiescent compared to the control sample. However, we not only find that the galaxy population differs, but also the dark matter halo population. 

The environmental difference between the infall and control region is revealed through the distributions of satellite galaxies around massive galaxies in both samples. Massive infall galaxies have twice as many satellite galaxies as similar mass control galaxies ($4.7\sigma$ significance). Both distributions are well fit by projected NFW profiles, but the haloes surrounding the massive infall galaxies have larger scale radii than the haloes around the control galaxies, even though the galaxies have the same stellar masses\footnote{The characteristic mass is the same in the protocluster and control samples and a Kolmogorov-Smirnov test finds no significant difference in the stellar mass distribution of galaxies with $M>10^{10.8}$\Msun\ in the two samples  ($p=0.97$)}.  Thus the massive infall galaxies are likely to be hosted by more massive haloes than similar mass galaxies from a control field. 

A small-scale density fluctuation collapses earlier if it lies within a region of large-scale overdensity such as a protocluster \citep{Kaiser_1984, Cole_1989, Mo_1996, Sheth2001}. This is the basis of biased clustering,  where more massive halos are more biased tracers of the underlying dark matter. Protoclusters are therefore expected to contain a relative abundance of massive collapsed objects, such as grouped-sized halos, compared to lower mass haloes, that host galaxies of typical or low masses, with respect to the field. Our measurement of the difference in halo properties between protocluster and control galaxies of similar mass allows us to directly observe this bias within protoclusters.

One important implication of this result is that the control sample is not an appropriate substitute for infall galaxies when calculating the efficiency and mass dependency of satellite  quenching in $z\sim1$ clusters. Using the infall population we calculate the QFE$_{cl-inf}$: the fraction of infalling galaxies that must be quenched once they fall into the cluster. Figure \ref{QFE} shows that $\sim40$\% of the star-forming infall galaxies must be quenched in the cluster and that there is no evidence of a mass dependence in the quenching efficiency over the mass range probed ($10^{10}<M/$\Msun$<10^{11.2}$). Above this mass, all infall and cluster masses are already quiescent so no further quenching is required when they become satellites.  
As the infall galaxies enter the denser environment of the cluster, slightly less than half of them have quenched by $z\sim1$ due to a cluster-specific process, which our constant QFE$_{cl-inf}$ suggests is independent of stellar mass.

Several recent studies have shown that the QFE at $z\sim1$ is mass-dependent \citep{Kawinwanichakij_2017, Papovich_2018, PintosCastro_2019,Vanderburg2020}. However, all of these studies calculate the QFE using the lowest density bin or an average/representative sample as their \lq field\rq\ sample. Indeed, when we use our control sample as the low density sample we also find a mass-dependent QFE$_{cl-con}$ (see Figure \ref{QFE_remco}), in agreement with these works. However, the galaxies within the control sample are not the direct progenitors of the cluster galaxies, so this mass-dependent QFE$_{cl-con}$ should not be used to infer the mass-dependency of the satellite quenching process within clusters.  Instead, QFE$_{cl-con}$ provides a measure of the galaxy quenching engendered by a combination of mass and environmental quenching processes that occurs in both the infall and cluster regions, as well as the consequence of the environmental dependence of the halo mass function. 

\begin{figure}
\centering
\includegraphics[width=1\columnwidth]{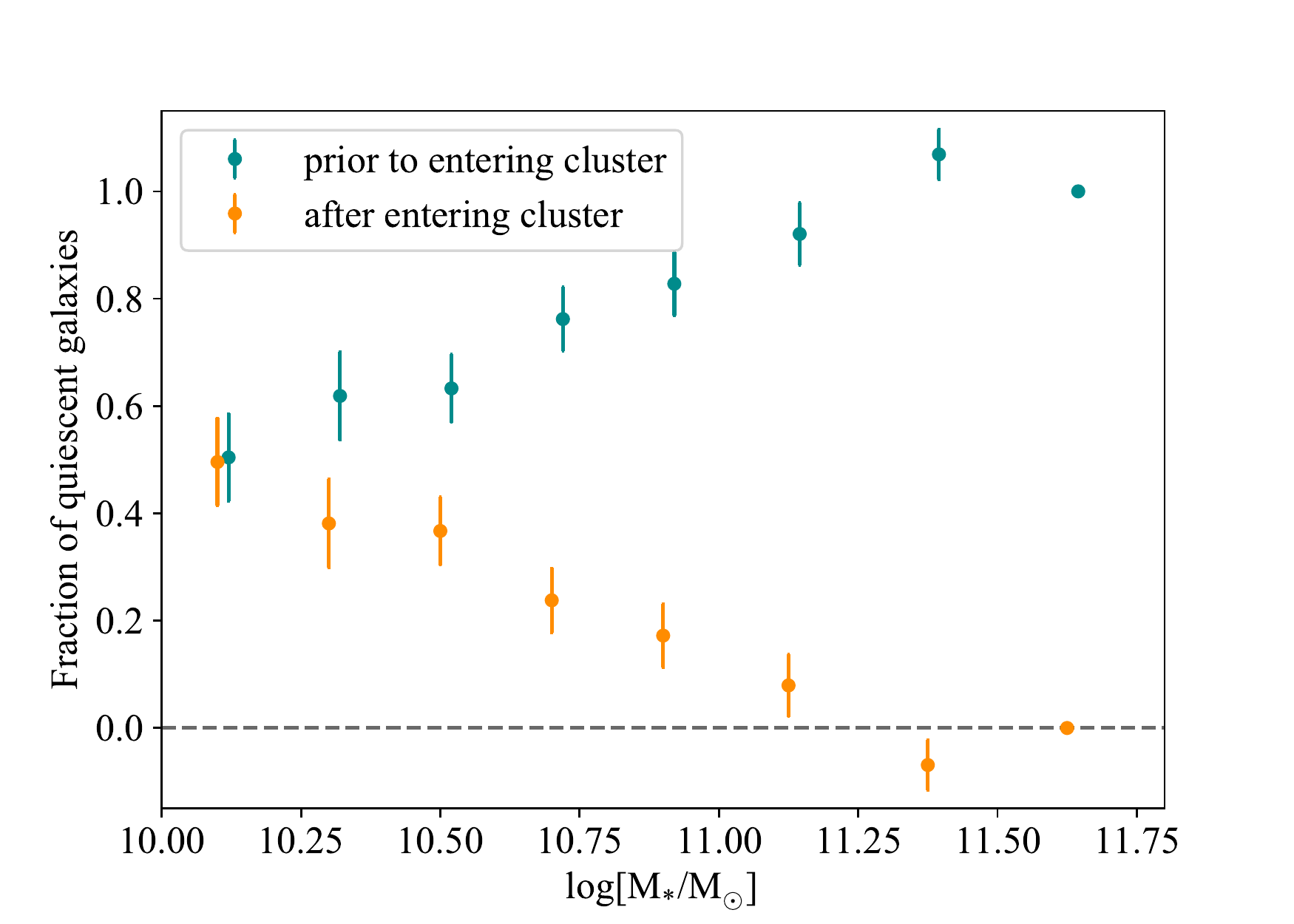}
\caption[parameters]{The fraction of quiescent cluster galaxies that were quenched {\it in} the cluster (blue) and {\it prior} to entering the cluster (orange).  Half of the lowest mass galaxies were quenched in the cluster, almost all of the massive cluster galaxies were quenched before they fell into the cluster.}
\label{fraction_quenched}
\end{figure}

A second implication of our results is that excess quenching of star formation occurs in the infall region relative to the control field. The QFE$_{inf-con}$ provides insight to the level and mass-dependency of the excess quenching that occurs in this infall/protocluster environment. In both the control and the infall sample we observe a strong mass dependence in the quiescent fraction (Fig.\,\ref{SMF}), which implies that the processes that quench galaxies in both environments are mass-dependent. But Figure \ref{QFE} shows that QFE$_{inf-con}$ has a strong dependency with stellar mass suggesting that the mass-quenching processes have an environmental dependence (in agreement with the results from \citealt{PintosCastro_2019}). 

We note that the stellar mass to halo mass relation is relatively flat above \Msun$>10^{10.5}$ \citep{Behroozi2015}, so a sample of central galaxies with a narrow range of stellar mass above this limit will inhabit a wide range of dark matter halo masses. If the infall region has a higher density of high mass halos or galaxy groups than the control region, as we hypothesise, then the same high stellar mass is picking up galaxies within different halo masses/environments in the two regions.

The excess of massive haloes in the infall region/protoclusters may enhance the quenching rate of galaxies. Contenders for this mass-dependant process include the ‘overconsumption’ model of \citet{McGee_2014} that affects satellite galaxies, or the ‘halo-quenching’ model of \citet{Dekel_Birnboim_2006} or ‘AGN-quenching’ models \citep[e.g.][]{Benson_2003, Granato_2004, Bower_2006, Croton_2006, Cattaneo_2009} that affects central galaxies. AGN-quenching is an attractive possibility since cosmological simulations of galaxy evolution show a correlation between black hole and halo mass \citep{Booth_2010} and observations suggest an enhancement of the AGN fraction within protoclusters \citep{Krishnan_2017}. However, with the limited data available we are unable to postulate which physical mechanisms are the most relevant for quenching the massive infalling galaxies.
 
Because of the excess quenching in the infall region, our results suggest that some cluster galaxies were quenched long before they entered the cluster. We illustrate this in Figure \ref{fraction_quenched} where we show the fraction of quiescent cluster galaxies that quenched {\it after} they fell into the cluster. This is calculated as $\frac{f_{{\rm q},cl} - f_{{\rm q},inf}}{f{_{\rm q,cl}}}$, where $f_{{\rm q},cl/inf}$ is the quiescent fraction in the cluster ($cl$) or infall ($inf$) region. We also plot the opposite of this fraction, which is the fraction of quiescent galaxies that were quenched {\it prior} to falling into the cluster. We find that almost all quiescent galaxies with $M>10^{11}$\Msun\ were quenched  prior to entering the cluster, whilst up to half of the lower mass galaxies were environmentally quenched after they passed the virial radius. This means the majority of massive quiescent cluster galaxies at $z\sim1$ were quenched not by satellite quenching in the cluster, but were self-quenched or pre-processed prior to infall. 

The hypothesis sketched by this paper agrees with the stellar age dating work of \citet{Webb_2020}, who measured the ages of field and GOGREEN cluster galaxies at $z\sim1$ and found that there is, at most, only a small age difference between them. Since we find that most cluster galaxies were not quenched in the cluster but in the infall region, their ages will reflect the long quenching timescales of self-quenched galaxies rather than the short timescales due to satellite quenching.

We conclude that environmental quenching of satellites within the main halo is not the most important quenching process within the highest redshift clusters, particularly for the most massive cluster members. We therefore surmise that the excess quiescent galaxies observed in these early clusters was a consequence of the protocluster environment in which these cluster galaxies formed and evolved.

\section{Summary and Conclusions}

We have analysed the galaxies in the outskirt regions of fifteen $0.8<z<1.4$ clusters from the GOGREEN and GCLASS surveys. We compared the masses and quiescent fractions of these infalling galaxies to cluster galaxies and a control sample of field galaxies. 

We find that infall galaxies differ significantly from the control and cluster galaxies in terms of their stellar mass distribution and quiescent fraction. The infall regions contain an excess of massive quiescent galaxies with respect to the control sample, and lack low-mass quiescent galaxies compared to the cluster sample. 
We find that massive infall galaxies are surrounded by twice as many satellites as control galaxies. These infalling satellites are distributed according to an NFW profile with a larger scale radius than the control field. Both of these results suggest that the dark matter halos surrounding infall galaxies are larger than those surrounding similar mass galaxies in the control field. We infer that the infall region contains a top-heavy halo mass function compared to the control. This different halo environment may be responsible for the excess quenching seen in the infall region compared to the control field via halo-quenching of central galaxies, or increased pre-processing of satellites.

We calculate the excess of quiescent galaxies caused by environmental quenching in the cluster by comparing the infalling galaxies to the cluster galaxies. We find that quenching of satellites in high-redshift clusters is independent of stellar mass. Furthermore, whilst satellite quenching is responsible for $\sim50$\% of the low-mass quenched galaxies in the cluster, almost all of the quiescent $>10^{11}$\Msun\ galaxies are quenched before entering the cluster. Thus, most of the excess quiescent galaxies present in high redshift clusters were quenched at an earlier phase when galaxies evolved in the protocluster/infall environment. 

We caveat that these results are limited to only $M>10^{10}$\Msun\ galaxies. Furthermore, the presence of backsplash galaxies and photometric redshift errors may impact our results in ways that are not apparent in the robustness checks we have performed.

\section*{Data Availability}
The data underlying this article are available in the GOGREEN survey site, at \url{http://gogreensurvey.ca/data-releases/data-packages/gogreen-and-gclass-first-data-release/}. 

\section*{Acknowledgements}
We sincerely thank Darren Croton for reviewing and helping us improve this paper.
SVW acknowledges support from the University of Nottingham through a Vice-Chancellor's International Scholarship. GW acknowledges support from the National Science Foundation through grant AST-1517863, by HST program number GO-15294,  and by grant number 80NSSC17K0019 issued through the NASA Astrophysics Data Analysis Program (ADAP). Support for program number GO-15294 was provided by NASA through a grant from the Space Telescope Science Institute, which is operated by the Association of Universities for Research in Astronomy, Incorporated, under NASA contract NAS5-26555.




\bibliographystyle{mnras}
\bibliography{Bibliography} 

\begin{thebibliography}{}
\makeatletter
\relax
\def\mn@urlcharsother{\let\do\@makeother \do\$\do\&\do\#\do\^\do\_\do\%\do\~}
\def\mn@doi{\begingroup\mn@urlcharsother \@ifnextchar [ {\mn@doi@}
  {\mn@doi@[]}}
\def\mn@doi@[#1]#2{\def\@tempa{#1}\ifx\@tempa\@empty \href
  {http://dx.doi.org/#2} {doi:#2}\else \href {http://dx.doi.org/#2} {#1}\fi
  \endgroup}
\def\mn@eprint#1#2{\mn@eprint@#1:#2::\@nil}
\def\mn@eprint@arXiv#1{\href {http://arxiv.org/abs/#1} {{\tt arXiv:#1}}}
\def\mn@eprint@dblp#1{\href {http://dblp.uni-trier.de/rec/bibtex/#1.xml}
  {dblp:#1}}
\def\mn@eprint@#1:#2:#3:#4\@nil{\def\@tempa {#1}\def\@tempb {#2}\def\@tempc
  {#3}\ifx \@tempc \@empty \let \@tempc \@tempb \let \@tempb \@tempa \fi \ifx
  \@tempb \@empty \def\@tempb {arXiv}\fi \@ifundefined
  {mn@eprint@\@tempb}{\@tempb:\@tempc}{\expandafter \expandafter \csname
  mn@eprint@\@tempb\endcsname \expandafter{\@tempc}}}

\bibitem[\protect\citeauthoryear{{Balogh}, {Baldry}, {Nichol}, {Miller},
  {Bower}  \& {Glazebrook}}{{Balogh} et~al.}{2004}]{Balogh2004}
{Balogh} M.~L.,  {Baldry} I.~K.,  {Nichol} R.,  {Miller} C.,  {Bower} R.,
  {Glazebrook} K.,  2004, \mn@doi [\apjl] {10.1086/426079}, \href
  {https://ui.adsabs.harvard.edu/abs/2004ApJ...615L.101B} {615, L101}

\bibitem[\protect\citeauthoryear{Balogh et~al.,}{Balogh
  et~al.}{2016}]{Balogh_2016}
Balogh M.~L.,  et~al., 2016, \mn@doi [Monthly Notices of the Royal Astronomical
  Society] {10.1093/mnras/stv2949}, 456, 4364–4376

\bibitem[\protect\citeauthoryear{{Balogh} et~al.,}{{Balogh}
  et~al.}{2017}]{Balogh_2017}
{Balogh} M.~L.,  et~al., 2017, \mn@doi [\mnras] {10.1093/mnras/stx1370}, \href
  {https://ui.adsabs.harvard.edu/abs/2017MNRAS.470.4168B} {470, 4168}

\bibitem[\protect\citeauthoryear{{Balogh} et~al.,}{{Balogh}
  et~al.}{2021}]{Balogh_2021}
{Balogh} M.~L.,  et~al., 2021, \mn@doi [\mnras] {10.1093/mnras/staa3008}, \href
  {https://ui.adsabs.harvard.edu/abs/2021MNRAS.500..358B} {500, 358}

\bibitem[\protect\citeauthoryear{{Bartelmann}}{{Bartelmann}}{1996}]{Bartelmann_1996}
{Bartelmann} M.,  1996, \aap, \href
  {https://ui.adsabs.harvard.edu/abs/1996A&A...313..697B} {313, 697}

\bibitem[\protect\citeauthoryear{{Behroozi} \& {Silk}}{{Behroozi} \&
  {Silk}}{2015}]{Behroozi2015}
{Behroozi} P.~S.,  {Silk} J.,  2015, \mn@doi [\apj]
  {10.1088/0004-637X/799/1/32}, \href
  {https://ui.adsabs.harvard.edu/abs/2015ApJ...799...32B} {799, 32}

\bibitem[\protect\citeauthoryear{{Benson}, {Bower}, {Frenk}, {Lacey}, {Baugh}
  \& {Cole}}{{Benson} et~al.}{2003}]{Benson_2003}
{Benson} A.~J.,  {Bower} R.~G.,  {Frenk} C.~S.,  {Lacey} C.~G.,  {Baugh} C.~M.,
    {Cole} S.,  2003, \mn@doi [\apj] {10.1086/379160}, \href
  {https://ui.adsabs.harvard.edu/abs/2003ApJ...599...38B} {599, 38}

\bibitem[\protect\citeauthoryear{{Bianconi}, {Smith}, {Haines}, {McGee},
  {Finoguenov}  \& {Egami}}{{Bianconi} et~al.}{2018}]{Bianconi2018}
{Bianconi} M.,  {Smith} G.~P.,  {Haines} C.~P.,  {McGee} S.~L.,  {Finoguenov}
  A.,   {Egami} E.,  2018, \mn@doi [\mnras] {10.1093/mnrasl/slx167}, \href
  {https://ui.adsabs.harvard.edu/abs/2018MNRAS.473L..79B} {473, L79}

\bibitem[\protect\citeauthoryear{Biviano et~al.,}{Biviano
  et~al.}{2021}]{Biviano_2021}
Biviano A.,  et~al., 2021, The GOGREEN survey: the internal dynamics of
  clusters of galaxies at redshift 0.9-1.4 (\mn@eprint {arXiv} {2104.01183})

\bibitem[\protect\citeauthoryear{Booth \& Schaye}{Booth \&
  Schaye}{2010}]{Booth_2010}
Booth C.~M.,  Schaye J.,  2010, \mn@doi [Monthly Notices of the Royal
  Astronomical Society: Letters] {10.1111/j.1745-3933.2010.00832.x}, 405,
  L1–L5

\bibitem[\protect\citeauthoryear{{Bower}, {Benson}, {Malbon}, {Helly}, {Frenk},
  {Baugh}, {Cole}  \& {Lacey}}{{Bower} et~al.}{2006}]{Bower_2006}
{Bower} R.~G.,  {Benson} A.~J.,  {Malbon} R.,  {Helly} J.~C.,  {Frenk} C.~S.,
  {Baugh} C.~M.,  {Cole} S.,   {Lacey} C.~G.,  2006, \mn@doi [\mnras]
  {10.1111/j.1365-2966.2006.10519.x}, \href
  {https://ui.adsabs.harvard.edu/abs/2006MNRAS.370..645B} {370, 645}

\bibitem[\protect\citeauthoryear{{Brammer}, {van Dokkum}  \& {Coppi}}{{Brammer}
  et~al.}{2008}]{Brammer_2008}
{Brammer} G.~B.,  {van Dokkum} P.~G.,   {Coppi} P.,  2008, \mn@doi [\apj]
  {10.1086/591786}, \href
  {https://ui.adsabs.harvard.edu/abs/2008ApJ...686.1503B} {686, 1503}

\bibitem[\protect\citeauthoryear{{Bruzual} \& {Charlot}}{{Bruzual} \&
  {Charlot}}{2003}]{Bruzual2003}
{Bruzual} G.,  {Charlot} S.,  2003, \mn@doi [\mnras]
  {10.1046/j.1365-8711.2003.06897.x}, \href
  {https://ui.adsabs.harvard.edu/abs/2003MNRAS.344.1000B} {344, 1000}

\bibitem[\protect\citeauthoryear{{Calzetti}, {Armus}, {Bohlin}, {Kinney},
  {Koornneef}  \& {Storchi-Bergmann}}{{Calzetti} et~al.}{2000}]{Calzetti2000}
{Calzetti} D.,  {Armus} L.,  {Bohlin} R.~C.,  {Kinney} A.~L.,  {Koornneef} J.,
   {Storchi-Bergmann} T.,  2000, \mn@doi [\apj] {10.1086/308692}, \href
  {https://ui.adsabs.harvard.edu/abs/2000ApJ...533..682C} {533, 682}

\bibitem[\protect\citeauthoryear{Cattaneo \& Best}{Cattaneo \&
  Best}{2009}]{Cattaneo_2009}
Cattaneo A.,  Best P.~N.,  2009, \mn@doi [Monthly Notices of the Royal
  Astronomical Society] {10.1111/j.1365-2966.2009.14557.x}, 395, 518

\bibitem[\protect\citeauthoryear{{Chabrier}}{{Chabrier}}{2003}]{Chabrier2003}
{Chabrier} G.,  2003, \mn@doi [\pasp] {10.1086/376392}, \href
  {https://ui.adsabs.harvard.edu/abs/2003PASP..115..763C} {115, 763}

\bibitem[\protect\citeauthoryear{{Chiang}, {Overzier}  \& {Gebhardt}}{{Chiang}
  et~al.}{2013}]{Chiang2013}
{Chiang} Y.-K.,  {Overzier} R.,   {Gebhardt} K.,  2013, \mn@doi [\apj]
  {10.1088/0004-637X/779/2/127}, \href
  {https://ui.adsabs.harvard.edu/abs/2013ApJ...779..127C} {779, 127}

\bibitem[\protect\citeauthoryear{{Chiang}, {Overzier}, {Gebhardt}  \&
  {Henriques}}{{Chiang} et~al.}{2017}]{Chiang2017}
{Chiang} Y.-K.,  {Overzier} R.~A.,  {Gebhardt} K.,   {Henriques} B.,  2017,
  \mn@doi [\apjl] {10.3847/2041-8213/aa7e7b}, \href
  {https://ui.adsabs.harvard.edu/abs/2017ApJ...844L..23C} {844, L23}

\bibitem[\protect\citeauthoryear{{Cole} \& {Kaiser}}{{Cole} \&
  {Kaiser}}{1989}]{Cole_1989}
{Cole} S.,  {Kaiser} N.,  1989, \mn@doi [\mnras] {10.1093/mnras/237.4.1127},
  \href {https://ui.adsabs.harvard.edu/abs/1989MNRAS.237.1127C} {237, 1127}

\bibitem[\protect\citeauthoryear{{Cooke}, {Hatch}, {Muldrew}, {Rigby}  \&
  {Kurk}}{{Cooke} et~al.}{2014}]{Cooke2014}
{Cooke} E.~A.,  {Hatch} N.~A.,  {Muldrew} S.~I.,  {Rigby} E.~E.,   {Kurk}
  J.~D.,  2014, \mn@doi [\mnras] {10.1093/mnras/stu522}, \href
  {https://ui.adsabs.harvard.edu/abs/2014MNRAS.440.3262C} {440, 3262}

\bibitem[\protect\citeauthoryear{{Cooke} et~al.,}{{Cooke}
  et~al.}{2016}]{Cook_2016}
{Cooke} E.~A.,  et~al., 2016, \mn@doi [\apj] {10.3847/0004-637X/816/2/83},
  \href {https://ui.adsabs.harvard.edu/abs/2016ApJ...816...83C} {816, 83}

\bibitem[\protect\citeauthoryear{{Cooper} et~al.,}{{Cooper}
  et~al.}{2010}]{Cooper2010}
{Cooper} M.~C.,  et~al., 2010, \mn@doi [\mnras]
  {10.1111/j.1365-2966.2010.17312.x}, \href
  {https://ui.adsabs.harvard.edu/abs/2010MNRAS.409..337C} {409, 337}

\bibitem[\protect\citeauthoryear{Croton et~al.,}{Croton
  et~al.}{2006}]{Croton_2006}
Croton D.~J.,  et~al., 2006, \mn@doi [Monthly Notices of the Royal Astronomical
  Society] {10.1111/j.1365-2966.2005.09675.x}, 365, 11–28

\bibitem[\protect\citeauthoryear{{De Lucia}, {Weinmann}, {Poggianti},
  {Arag{\'o}n-Salamanca}  \& {Zaritsky}}{{De Lucia} et~al.}{2012}]{DeLucia2012}
{De Lucia} G.,  {Weinmann} S.,  {Poggianti} B.~M.,  {Arag{\'o}n-Salamanca} A.,
   {Zaritsky} D.,  2012, \mn@doi [\mnras] {10.1111/j.1365-2966.2012.20983.x},
  \href {https://ui.adsabs.harvard.edu/abs/2012MNRAS.423.1277D} {423, 1277}

\bibitem[\protect\citeauthoryear{{Dekel} \& {Birnboim}}{{Dekel} \&
  {Birnboim}}{2006}]{Dekel_Birnboim_2006}
{Dekel} A.,  {Birnboim} Y.,  2006, \mn@doi [\mnras]
  {10.1111/j.1365-2966.2006.10145.x}, \href
  {https://ui.adsabs.harvard.edu/abs/2006MNRAS.368....2D} {368, 2}

\bibitem[\protect\citeauthoryear{{Diemer} \& {Kravtsov}}{{Diemer} \&
  {Kravtsov}}{2014}]{DiemerKravtsov_2014}
{Diemer} B.,  {Kravtsov} A.~V.,  2014, \mn@doi [\apj]
  {10.1088/0004-637X/789/1/1}, \href
  {https://ui.adsabs.harvard.edu/abs/2014ApJ...789....1D} {789, 1}

\bibitem[\protect\citeauthoryear{{Dressler}}{{Dressler}}{1980}]{Dressler1980}
{Dressler} A.,  1980, \mn@doi [\apj] {10.1086/157753}, \href
  {https://ui.adsabs.harvard.edu/abs/1980ApJ...236..351D} {236, 351}

\bibitem[\protect\citeauthoryear{{Dressler} et~al.,}{{Dressler}
  et~al.}{1997}]{Dressler1997}
{Dressler} A.,  et~al., 1997, \mn@doi [\apj] {10.1086/304890}, \href
  {https://ui.adsabs.harvard.edu/abs/1997ApJ...490..577D} {490, 577}

\bibitem[\protect\citeauthoryear{{Finoguenov} et~al.,}{{Finoguenov}
  et~al.}{2007}]{Finoguenov_2007}
{Finoguenov} A.,  et~al., 2007, \mn@doi [\apjs] {10.1086/516577}, \href
  {https://ui.adsabs.harvard.edu/abs/2007ApJS..172..182F} {172, 182}

\bibitem[\protect\citeauthoryear{{Finoguenov} et~al.,}{{Finoguenov}
  et~al.}{2010}]{Finoguenov_2010}
{Finoguenov} A.,  et~al., 2010, \mn@doi [\mnras]
  {10.1111/j.1365-2966.2010.16256.x}, \href
  {https://ui.adsabs.harvard.edu/abs/2010MNRAS.403.2063F} {403, 2063}

\bibitem[\protect\citeauthoryear{{Fioc} \& {Rocca-Volmerange}}{{Fioc} \&
  {Rocca-Volmerange}}{1997}]{Fioc1997}
{Fioc} M.,  {Rocca-Volmerange} B.,  1997, \aap, \href
  {https://ui.adsabs.harvard.edu/abs/1997A&A...326..950F} {500, 507}

\bibitem[\protect\citeauthoryear{Foley et~al.,}{Foley
  et~al.}{2011}]{Foley_2011}
Foley R.~J.,  et~al., 2011, \mn@doi [The Astrophysical Journal]
  {10.1088/0004-637x/731/2/86}, 731, 86

\bibitem[\protect\citeauthoryear{{Foltz} et~al.,}{{Foltz}
  et~al.}{2018}]{Foltz_2018}
{Foltz} R.,  et~al., 2018, \mn@doi [\apj] {10.3847/1538-4357/aad80d}, \href
  {https://ui.adsabs.harvard.edu/abs/2018ApJ...866..136F} {866, 136}

\bibitem[\protect\citeauthoryear{Foreman-Mackey, Hogg, Lang  \&
  Goodman}{Foreman-Mackey et~al.}{2013}]{emcee2012}
Foreman-Mackey D.,  Hogg D.~W.,  Lang D.,   Goodman J.,  2013, \mn@doi
  [Publications of the Astronomical Society of the Pacific] {10.1086/670067},
  125, 306–312

\bibitem[\protect\citeauthoryear{Fossati et~al.,}{Fossati
  et~al.}{2017}]{Fossati_2017}
Fossati M.,  et~al., 2017, \mn@doi [The Astrophysical Journal]
  {10.3847/1538-4357/835/2/153}, 835, 153

\bibitem[\protect\citeauthoryear{{George} et~al.,}{{George}
  et~al.}{2011}]{George_2011}
{George} M.~R.,  et~al., 2011, \mn@doi [\apj] {10.1088/0004-637X/742/2/125},
  \href {https://ui.adsabs.harvard.edu/abs/2011ApJ...742..125G} {742, 125}

\bibitem[\protect\citeauthoryear{{Giodini} et~al.,}{{Giodini}
  et~al.}{2012}]{Giodini2012}
{Giodini} S.,  et~al., 2012, \mn@doi [\aap] {10.1051/0004-6361/201117696},
  \href {https://ui.adsabs.harvard.edu/abs/2012A&A...538A.104G} {538, A104}

\bibitem[\protect\citeauthoryear{{Gobat}, {Rosati}, {Strazzullo}, {Rettura},
  {Demarco}  \& {Nonino}}{{Gobat} et~al.}{2008}]{Gobat2008}
{Gobat} R.,  {Rosati} P.,  {Strazzullo} V.,  {Rettura} A.,  {Demarco} R.,
  {Nonino} M.,  2008, \mn@doi [\aap] {10.1051/0004-6361:200809531}, \href
  {https://ui.adsabs.harvard.edu/abs/2008A&A...488..853G} {488, 853}

\bibitem[\protect\citeauthoryear{{G{\'o}mez} et~al.,}{{G{\'o}mez}
  et~al.}{2003}]{Gomez2003}
{G{\'o}mez} P.~L.,  et~al., 2003, \mn@doi [\apj] {10.1086/345593}, \href
  {https://ui.adsabs.harvard.edu/abs/2003ApJ...584..210G} {584, 210}

\bibitem[\protect\citeauthoryear{{Granato}, {De Zotti}, {Silva}, {Bressan}  \&
  {Danese}}{{Granato} et~al.}{2004}]{Granato_2004}
{Granato} G.~L.,  {De Zotti} G.,  {Silva} L.,  {Bressan} A.,   {Danese} L.,
  2004, \mn@doi [\apj] {10.1086/379875}, \href
  {https://ui.adsabs.harvard.edu/abs/2004ApJ...600..580G} {600, 580}

\bibitem[\protect\citeauthoryear{{Gunn} \& {Gott}}{{Gunn} \&
  {Gott}}{1972}]{Gunn1972}
{Gunn} J.~E.,  {Gott} J.~Richard I.,  1972, \mn@doi [\apj] {10.1086/151605},
  \href {https://ui.adsabs.harvard.edu/abs/1972ApJ...176....1G} {176, 1}

\bibitem[\protect\citeauthoryear{{Haggar}, {Gray}, {Pearce}, {Knebe}, {Cui},
  {Mostoghiu}  \& {Yepes}}{{Haggar} et~al.}{2020}]{Haggar_2020}
{Haggar} R.,  {Gray} M.~E.,  {Pearce} F.~R.,  {Knebe} A.,  {Cui} W.,
  {Mostoghiu} R.,   {Yepes} G.,  2020, \mn@doi [\mnras]
  {10.1093/mnras/staa273}, \href
  {https://ui.adsabs.harvard.edu/abs/2020MNRAS.492.6074H} {492, 6074}

\bibitem[\protect\citeauthoryear{Haines et~al.,}{Haines
  et~al.}{2015}]{Haines2015}
Haines C.~P.,  et~al., 2015, \mn@doi [The Astrophysical Journal]
  {10.1088/0004-637x/806/1/101}, 806, 101

\bibitem[\protect\citeauthoryear{{Hatch}, {Kurk}, {Pentericci}, {Venemans},
  {Kuiper}, {Miley}  \& {R{\"o}ttgering}}{{Hatch} et~al.}{2011}]{Hatch2011}
{Hatch} N.~A.,  {Kurk} J.~D.,  {Pentericci} L.,  {Venemans} B.~P.,  {Kuiper}
  E.,  {Miley} G.~K.,   {R{\"o}ttgering} H.~J.~A.,  2011, \mn@doi [\mnras]
  {10.1111/j.1365-2966.2011.18735.x}, \href
  {https://ui.adsabs.harvard.edu/abs/2011MNRAS.415.2993H} {415, 2993}

\bibitem[\protect\citeauthoryear{Just et~al.,}{Just et~al.}{2019}]{Just2019}
Just D.~W.,  et~al., 2019, \mn@doi [The Astrophysical Journal]
  {10.3847/1538-4357/ab44a0}, 885, 6

\bibitem[\protect\citeauthoryear{{Kaiser}}{{Kaiser}}{1984}]{Kaiser_1984}
{Kaiser} N.,  1984, \mn@doi [\apjl] {10.1086/184341}, \href
  {https://ui.adsabs.harvard.edu/abs/1984ApJ...284L...9K} {284, L9}

\bibitem[\protect\citeauthoryear{Kawinwanichakij et~al.,}{Kawinwanichakij
  et~al.}{2014}]{Kawinwanichakij_2014}
Kawinwanichakij L.,  et~al., 2014, \mn@doi [The Astrophysical Journal]
  {10.1088/0004-637x/792/2/103}, 792, 103

\bibitem[\protect\citeauthoryear{Kawinwanichakij et~al.,}{Kawinwanichakij
  et~al.}{2017}]{Kawinwanichakij_2017}
Kawinwanichakij L.,  et~al., 2017, \mn@doi [The Astrophysical Journal]
  {10.3847/1538-4357/aa8b75}, 847, 134

\bibitem[\protect\citeauthoryear{{Kodama}, {Balogh}, {Smail}, {Bower}  \&
  {Nakata}}{{Kodama} et~al.}{2004}]{Kodama2004}
{Kodama} T.,  {Balogh} M.~L.,  {Smail} I.,  {Bower} R.~G.,   {Nakata} F.,
  2004, \mn@doi [\mnras] {10.1111/j.1365-2966.2004.08271.x}, \href
  {https://ui.adsabs.harvard.edu/abs/2004MNRAS.354.1103K} {354, 1103}

\bibitem[\protect\citeauthoryear{{Koyama}, {Kodama}, {Nakata}, {Shimasaku}  \&
  {Okamura}}{{Koyama} et~al.}{2011}]{Koyama2011}
{Koyama} Y.,  {Kodama} T.,  {Nakata} F.,  {Shimasaku} K.,   {Okamura} S.,
  2011, \mn@doi [\apj] {10.1088/0004-637X/734/1/66}, \href
  {https://ui.adsabs.harvard.edu/abs/2011ApJ...734...66K} {734, 66}

\bibitem[\protect\citeauthoryear{{Kriek} et~al.,}{{Kriek}
  et~al.}{2018}]{Kriek2018}
{Kriek} M.,  et~al., 2018, {FAST: Fitting and Assessment of Synthetic
  Templates} (\mn@eprint {ascl} {1803.008})

\bibitem[\protect\citeauthoryear{Krishnan et~al.,}{Krishnan
  et~al.}{2017}]{Krishnan_2017}
Krishnan C.,  et~al., 2017, \mn@doi [Monthly Notices of the Royal Astronomical
  Society] {10.1093/mnras/stx1315}, 470, 2170–2178

\bibitem[\protect\citeauthoryear{{Lee-Brown} et~al.,}{{Lee-Brown}
  et~al.}{2017}]{Lee-Brown2017}
{Lee-Brown} D.~B.,  et~al., 2017, \mn@doi [\apj] {10.3847/1538-4357/aa7948},
  \href {https://ui.adsabs.harvard.edu/abs/2017ApJ...844...43L} {844, 43}

\bibitem[\protect\citeauthoryear{{Leja}, {Carnall}, {Johnson}, {Conroy}  \&
  {Speagle}}{{Leja} et~al.}{2019}]{Leja_2019}
{Leja} J.,  {Carnall} A.~C.,  {Johnson} B.~D.,  {Conroy} C.,   {Speagle} J.~S.,
   2019, \mn@doi [\apj] {10.3847/1538-4357/ab133c}, \href
  {https://ui.adsabs.harvard.edu/abs/2019ApJ...876....3L} {876, 3}

\bibitem[\protect\citeauthoryear{{Lewis} et~al.,}{{Lewis}
  et~al.}{2002}]{Lewis2002}
{Lewis} I.,  et~al., 2002, \mn@doi [\mnras] {10.1046/j.1365-8711.2002.05558.x},
  \href {https://ui.adsabs.harvard.edu/abs/2002MNRAS.334..673L} {334, 673}

\bibitem[\protect\citeauthoryear{{Maraston}}{{Maraston}}{2005}]{Maraston2005}
{Maraston} C.,  2005, \mn@doi [\mnras] {10.1111/j.1365-2966.2005.09270.x},
  \href {https://ui.adsabs.harvard.edu/abs/2005MNRAS.362..799M} {362, 799}

\bibitem[\protect\citeauthoryear{{McGee}, {Bower}  \& {Balogh}}{{McGee}
  et~al.}{2014}]{McGee_2014}
{McGee} S.~L.,  {Bower} R.~G.,   {Balogh} M.~L.,  2014, \mn@doi [\mnras]
  {10.1093/mnrasl/slu066}, \href
  {https://ui.adsabs.harvard.edu/abs/2014MNRAS.442L.105M} {442, L105}

\bibitem[\protect\citeauthoryear{{Mo}, {Jing}  \& {White}}{{Mo}
  et~al.}{1996}]{Mo_1996}
{Mo} H.~J.,  {Jing} Y.~P.,   {White} S.~D.~M.,  1996, \mn@doi [\mnras]
  {10.1093/mnras/282.3.1096}, \href
  {https://ui.adsabs.harvard.edu/abs/1996MNRAS.282.1096M} {282, 1096}

\bibitem[\protect\citeauthoryear{{Moore}, {Katz}, {Lake}, {Dressler}  \&
  {Oemler}}{{Moore} et~al.}{1996}]{Moore1996}
{Moore} B.,  {Katz} N.,  {Lake} G.,  {Dressler} A.,   {Oemler} A.,  1996,
  \mn@doi [\nat] {10.1038/379613a0}, \href
  {https://ui.adsabs.harvard.edu/abs/1996Natur.379..613M} {379, 613}

\bibitem[\protect\citeauthoryear{{Muldrew}, {Hatch}  \& {Cooke}}{{Muldrew}
  et~al.}{2015}]{Muldrew2015}
{Muldrew} S.~I.,  {Hatch} N.~A.,   {Cooke} E.~A.,  2015, \mn@doi [\mnras]
  {10.1093/mnras/stv1449}, \href
  {https://ui.adsabs.harvard.edu/abs/2015MNRAS.452.2528M} {452, 2528}

\bibitem[\protect\citeauthoryear{{Muldrew}, {Hatch}  \& {Cooke}}{{Muldrew}
  et~al.}{2018}]{Muldrew_2018}
{Muldrew} S.~I.,  {Hatch} N.~A.,   {Cooke} E.~A.,  2018, \mn@doi [\mnras]
  {10.1093/mnras/stx2454}, \href
  {https://ui.adsabs.harvard.edu/abs/2018MNRAS.473.2335M} {473, 2335}

\bibitem[\protect\citeauthoryear{{Muzzin} et~al.,}{{Muzzin}
  et~al.}{2009}]{Muzzin2009}
{Muzzin} A.,  et~al., 2009, \mn@doi [\apj] {10.1088/0004-637X/698/2/1934},
  \href {https://ui.adsabs.harvard.edu/abs/2009ApJ...698.1934M} {698, 1934}

\bibitem[\protect\citeauthoryear{{Muzzin} et~al.,}{{Muzzin}
  et~al.}{2012}]{Muzzin2012gclass}
{Muzzin} A.,  et~al., 2012, \mn@doi [\apj] {10.1088/0004-637X/746/2/188}, \href
  {https://ui.adsabs.harvard.edu/abs/2012ApJ...746..188M} {746, 188}

\bibitem[\protect\citeauthoryear{{Muzzin} et~al.,}{{Muzzin}
  et~al.}{2013}]{Muzzin2013}
{Muzzin} A.,  et~al., 2013, \mn@doi [\apj] {10.1088/0004-637X/777/1/18}, \href
  {https://ui.adsabs.harvard.edu/abs/2013ApJ...777...18M} {777, 18}

\bibitem[\protect\citeauthoryear{{Nantais} et~al.,}{{Nantais}
  et~al.}{2016}]{Nantais2016}
{Nantais} J.~B.,  et~al., 2016, \mn@doi [\aap] {10.1051/0004-6361/201628663},
  \href {https://ui.adsabs.harvard.edu/abs/2016A&A...592A.161N} {592, A161}

\bibitem[\protect\citeauthoryear{{Navarro}, {Frenk}  \& {White}}{{Navarro}
  et~al.}{1996}]{NFW_1996}
{Navarro} J.~F.,  {Frenk} C.~S.,   {White} S. D.~M.,  1996, \mn@doi [\apj]
  {10.1086/177173}, \href
  {https://ui.adsabs.harvard.edu/abs/1996ApJ...462..563N} {462, 563}

\bibitem[\protect\citeauthoryear{{Newman}, {Ellis}, {Andreon}, {Treu},
  {Raichoor}  \& {Trinchieri}}{{Newman} et~al.}{2014}]{Newman_2014}
{Newman} A.~B.,  {Ellis} R.~S.,  {Andreon} S.,  {Treu} T.,  {Raichoor} A.,
  {Trinchieri} G.,  2014, \mn@doi [\apj] {10.1088/0004-637X/788/1/51}, \href
  {https://ui.adsabs.harvard.edu/abs/2014ApJ...788...51N} {788, 51}

\bibitem[\protect\citeauthoryear{{Nierenberg}, {Auger}, {Treu}, {Marshall}  \&
  {Fassnacht}}{{Nierenberg} et~al.}{2011}]{Nierenberg_2011}
{Nierenberg} A.~M.,  {Auger} M.~W.,  {Treu} T.,  {Marshall} P.~J.,
  {Fassnacht} C.~D.,  2011, \mn@doi [\apj] {10.1088/0004-637X/731/1/44}, \href
  {https://ui.adsabs.harvard.edu/abs/2011ApJ...731...44N} {731, 44}

\bibitem[\protect\citeauthoryear{{Nierenberg}, {Auger}, {Treu}, {Marshall},
  {Fassnacht}  \& {Busha}}{{Nierenberg} et~al.}{2012}]{Nierenberg_2012}
{Nierenberg} A.~M.,  {Auger} M.~W.,  {Treu} T.,  {Marshall} P.~J.,  {Fassnacht}
  C.~D.,   {Busha} M.~T.,  2012, \mn@doi [\apj] {10.1088/0004-637X/752/2/99},
  \href {https://ui.adsabs.harvard.edu/abs/2012ApJ...752...99N} {752, 99}

\bibitem[\protect\citeauthoryear{{Oemler}, {Dressler}, {Gladders}, {Fritz},
  {Poggianti}, {Vulcani}  \& {Abramson}}{{Oemler} et~al.}{2013}]{Oemler2013}
{Oemler} Augustus J.,  {Dressler} A.,  {Gladders} M.~G.,  {Fritz} J.,
  {Poggianti} B.~M.,  {Vulcani} B.,   {Abramson} L.,  2013, \mn@doi [\apj]
  {10.1088/0004-637X/770/1/63}, \href
  {https://ui.adsabs.harvard.edu/abs/2013ApJ...770...63O} {770, 63}

\bibitem[\protect\citeauthoryear{{Old} et~al.,}{{Old} et~al.}{2021}]{Old2021}
{Old} L.~J.,  et~al., 2021, \mn@doi [\mnras] {10.1093/mnras/staa2890}, \href
  {https://ui.adsabs.harvard.edu/abs/2021MNRAS.500..355O} {500, 355}

\bibitem[\protect\citeauthoryear{{Papovich} et~al.,}{{Papovich}
  et~al.}{2016}]{Papovich2016}
{Papovich} C.,  et~al., 2016, \mn@doi [\apjs] {10.3847/0067-0049/224/2/28},
  \href {https://ui.adsabs.harvard.edu/abs/2016ApJS..224...28P} {224, 28}

\bibitem[\protect\citeauthoryear{{Papovich} et~al.,}{{Papovich}
  et~al.}{2018}]{Papovich_2018}
{Papovich} C.,  et~al., 2018, \mn@doi [\apj] {10.3847/1538-4357/aaa766}, \href
  {https://ui.adsabs.harvard.edu/abs/2018ApJ...854...30P} {854, 30}

\bibitem[\protect\citeauthoryear{{Patel}, {Kelson}, {Holden}, {Franx}  \&
  {Illingworth}}{{Patel} et~al.}{2011}]{Patel2011}
{Patel} S.~G.,  {Kelson} D.~D.,  {Holden} B.~P.,  {Franx} M.,   {Illingworth}
  G.~D.,  2011, \mn@doi [\apj] {10.1088/0004-637X/735/1/53}, \href
  {https://ui.adsabs.harvard.edu/abs/2011ApJ...735...53P} {735, 53}

\bibitem[\protect\citeauthoryear{{Peng} et~al.,}{{Peng}
  et~al.}{2010}]{Peng2010}
{Peng} Y.-j.,  et~al., 2010, \mn@doi [\apj] {10.1088/0004-637X/721/1/193},
  \href {https://ui.adsabs.harvard.edu/abs/2010ApJ...721..193P} {721, 193}

\bibitem[\protect\citeauthoryear{Pintos-Castro, Yee, Muzzin, Old  \&
  Wilson}{Pintos-Castro et~al.}{2019}]{PintosCastro_2019}
Pintos-Castro I.,  Yee H. K.~C.,  Muzzin A.,  Old L.,   Wilson G.,  2019,
  \mn@doi [The Astrophysical Journal] {10.3847/1538-4357/ab14ee}, 876, 40

\bibitem[\protect\citeauthoryear{{Rettura} et~al.,}{{Rettura}
  et~al.}{2010}]{Rettura2010}
{Rettura} A.,  et~al., 2010, \mn@doi [\apj] {10.1088/0004-637X/709/1/512},
  \href {https://ui.adsabs.harvard.edu/abs/2010ApJ...709..512R} {709, 512}

\bibitem[\protect\citeauthoryear{{Rudnick}, {Tran}, {Papovich}, {Momcheva}  \&
  {Willmer}}{{Rudnick} et~al.}{2012}]{Rudnick_2012}
{Rudnick} G.~H.,  {Tran} K.-V.,  {Papovich} C.,  {Momcheva} I.,   {Willmer} C.,
   2012, \mn@doi [\apj] {10.1088/0004-637X/755/1/14}, \href
  {https://ui.adsabs.harvard.edu/abs/2012ApJ...755...14R} {755, 14}

\bibitem[\protect\citeauthoryear{{Schechter}}{{Schechter}}{1976}]{Schechter1976}
{Schechter} P.,  1976, \mn@doi [\apj] {10.1086/154079}, \href
  {https://ui.adsabs.harvard.edu/abs/1976ApJ...203..297S} {203, 297}

\bibitem[\protect\citeauthoryear{Schneider}{Schneider}{2006}]{Schneider2006}
Schneider P.,  2006, Extragalactic Astronomy and Cosmology.
Springer Berlin Heidelberg, Berlin, Heidelberg,
  \mn@doi{10.1007/978-3-540-33175-9}, \url
  {http://dx.doi.org/10.1007/978-3-540-33175-9}

\bibitem[\protect\citeauthoryear{{Sheth}, {Mo}  \& {Tormen}}{{Sheth}
  et~al.}{2001}]{Sheth2001}
{Sheth} R.~K.,  {Mo} H.~J.,   {Tormen} G.,  2001, \mn@doi [\mnras]
  {10.1046/j.1365-8711.2001.04006.x}, \href
  {https://ui.adsabs.harvard.edu/abs/2001MNRAS.323....1S} {323, 1}

\bibitem[\protect\citeauthoryear{{Sif{\'o}n} et~al.,}{{Sif{\'o}n}
  et~al.}{2016}]{Sifon_2016}
{Sif{\'o}n} C.,  et~al., 2016, \mn@doi [\mnras] {10.1093/mnras/stw1284}, \href
  {https://ui.adsabs.harvard.edu/abs/2016MNRAS.461..248S} {461, 248}

\bibitem[\protect\citeauthoryear{{Sobral}, {Best}, {Smail}, {Geach},
  {Cirasuolo}, {Garn}  \& {Dalton}}{{Sobral} et~al.}{2011}]{Sobral2011}
{Sobral} D.,  {Best} P.~N.,  {Smail} I.,  {Geach} J.~E.,  {Cirasuolo} M.,
  {Garn} T.,   {Dalton} G.~B.,  2011, \mn@doi [\mnras]
  {10.1111/j.1365-2966.2010.17707.x}, \href
  {https://ui.adsabs.harvard.edu/abs/2011MNRAS.411..675S} {411, 675}

\bibitem[\protect\citeauthoryear{Stalder et~al.,}{Stalder
  et~al.}{2013}]{Stalder_2013}
Stalder B.,  et~al., 2013, \mn@doi [The Astrophysical Journal]
  {10.1088/0004-637x/763/2/93}, 763, 93

\bibitem[\protect\citeauthoryear{{Strazzullo} et~al.,}{{Strazzullo}
  et~al.}{2019}]{Strazzullo2019}
{Strazzullo} V.,  et~al., 2019, \mn@doi [\aap] {10.1051/0004-6361/201833944},
  \href {https://ui.adsabs.harvard.edu/abs/2019A&A...622A.117S} {622, A117}

\bibitem[\protect\citeauthoryear{{Sunyaev} \& {Zeldovich}}{{Sunyaev} \&
  {Zeldovich}}{1970}]{Sunyaev_Zeldovich1970}
{Sunyaev} R.~A.,  {Zeldovich} Y.~B.,  1970, Comments on Astrophysics and Space
  Physics, \href {https://ui.adsabs.harvard.edu/abs/1970CoASP...2...66S} {2,
  66}

\bibitem[\protect\citeauthoryear{{Tal}, {Wake}  \& {van Dokkum}}{{Tal}
  et~al.}{2012}]{Tal_2012}
{Tal} T.,  {Wake} D.~A.,   {van Dokkum} P.~G.,  2012, \mn@doi [\apjl]
  {10.1088/2041-8205/751/1/L5}, \href
  {https://ui.adsabs.harvard.edu/abs/2012ApJ...751L...5T} {751, L5}

\bibitem[\protect\citeauthoryear{{Tal}, {van Dokkum}, {Franx}, {Leja}, {Wake}
  \& {Whitaker}}{{Tal} et~al.}{2013}]{Tal_2013}
{Tal} T.,  {van Dokkum} P.~G.,  {Franx} M.,  {Leja} J.,  {Wake} D.~A.,
  {Whitaker} K.~E.,  2013, \mn@doi [\apj] {10.1088/0004-637X/769/1/31}, \href
  {https://ui.adsabs.harvard.edu/abs/2013ApJ...769...31T} {769, 31}

\bibitem[\protect\citeauthoryear{{Tanaka}, {Kodama}, {Arimoto}  \&
  {Tanaka}}{{Tanaka} et~al.}{2006}]{Tanaka2006}
{Tanaka} M.,  {Kodama} T.,  {Arimoto} N.,   {Tanaka} I.,  2006, \mn@doi
  [\mnras] {10.1111/j.1365-2966.2005.09841.x}, \href
  {https://ui.adsabs.harvard.edu/abs/2006MNRAS.365.1392T} {365, 1392}

\bibitem[\protect\citeauthoryear{Wang \& White}{Wang \&
  White}{2012}]{Wang_2012}
Wang W.,  White S. D.~M.,  2012, \mn@doi [Monthly Notices of the Royal
  Astronomical Society] {10.1111/j.1365-2966.2012.21256.x}, 424, 2574–2598

\bibitem[\protect\citeauthoryear{Wang, Sales, Henriques  \& White}{Wang
  et~al.}{2014}]{Wang_2014}
Wang W.,  Sales L.~V.,  Henriques B. M.~B.,   White S. D.~M.,  2014, \mn@doi
  [Monthly Notices of the Royal Astronomical Society] {10.1093/mnras/stu988},
  442, 1363–1378

\bibitem[\protect\citeauthoryear{{Watson}, {Berlind}  \& {Zentner}}{{Watson}
  et~al.}{2012}]{Watson_2012}
{Watson} D.~F.,  {Berlind} A.~A.,   {Zentner} A.~R.,  2012, \mn@doi [\apj]
  {10.1088/0004-637X/754/2/90}, \href
  {https://ui.adsabs.harvard.edu/abs/2012ApJ...754...90W} {754, 90}

\bibitem[\protect\citeauthoryear{{Webb} et~al.,}{{Webb}
  et~al.}{2020}]{Webb_2020}
{Webb} K.,  et~al., 2020, \mn@doi [\mnras] {10.1093/mnras/staa2752}, \href
  {https://ui.adsabs.harvard.edu/abs/2020MNRAS.498.5317W} {498, 5317}

\bibitem[\protect\citeauthoryear{{Wetzel}, {Tinker}, {Conroy}  \& {van den
  Bosch}}{{Wetzel} et~al.}{2013}]{Wetzel2013}
{Wetzel} A.~R.,  {Tinker} J.~L.,  {Conroy} C.,   {van den Bosch} F.~C.,  2013,
  \mn@doi [\mnras] {10.1093/mnras/stt469}, \href
  {https://ui.adsabs.harvard.edu/abs/2013MNRAS.432..336W} {432, 336}

\bibitem[\protect\citeauthoryear{Williams, Quadri, Franx, van Dokkum  \&
  Labbé}{Williams et~al.}{2009}]{Williams2009}
Williams R.~J.,  Quadri R.~F.,  Franx M.,  van Dokkum P.,   Labbé I.,  2009,
  \mn@doi [The Astrophysical Journal] {10.1088/0004-637x/691/2/1879}, 691,
  1879–1895

\bibitem[\protect\citeauthoryear{{Willis} et~al.,}{{Willis}
  et~al.}{2020}]{Willis2020}
{Willis} J.~P.,  et~al., 2020, \mn@doi [\nat] {10.1038/s41586-019-1829-4},
  \href {https://ui.adsabs.harvard.edu/abs/2020Natur.577...39W} {577, 39}

\bibitem[\protect\citeauthoryear{{Wilson}, {Muzzin}, {Yee}  \& {SpARCS
  Collaboration}}{{Wilson} et~al.}{2009a}]{Wilson20092}
{Wilson} G.,  {Muzzin} A.,  {Yee} H.,   {SpARCS Collaboration} 2009a, in
  American Astronomical Society Meeting Abstracts \#213. p. 315.01

\bibitem[\protect\citeauthoryear{{Wilson} et~al.,}{{Wilson}
  et~al.}{2009b}]{Wilson2009}
{Wilson} G.,  et~al., 2009b, \mn@doi [\apj] {10.1088/0004-637X/698/2/1943},
  \href {https://ui.adsabs.harvard.edu/abs/2009ApJ...698.1943W} {698, 1943}

\bibitem[\protect\citeauthoryear{van~den Bosch, Aquino, Yang, Mo, Pasquali,
  McIntosh, Weinmann  \& Kang}{van~den Bosch et~al.}{2008}]{vandenBosch_2008}
van~den Bosch F.~C.,  Aquino D.,  Yang X.,  Mo H.~J.,  Pasquali A.,  McIntosh
  D.~H.,  Weinmann S.~M.,   Kang X.,  2008, \mn@doi [Monthly Notices of the
  Royal Astronomical Society] {10.1111/j.1365-2966.2008.13230.x}, 387, 79–91

\bibitem[\protect\citeauthoryear{{van der Burg} et~al.,}{{van der Burg}
  et~al.}{2013}]{Vanderburg_2013}
{van der Burg} R.~F.~J.,  et~al., 2013, \mn@doi [\aap]
  {10.1051/0004-6361/201321237}, \href
  {https://ui.adsabs.harvard.edu/abs/2013A&A...557A..15V} {557, A15}

\bibitem[\protect\citeauthoryear{{van der Burg} et~al.,}{{van der Burg}
  et~al.}{2020}]{Vanderburg2020}
{van der Burg} R. F.~J.,  et~al., 2020, \mn@doi [\aap]
  {10.1051/0004-6361/202037754}, \href
  {https://ui.adsabs.harvard.edu/abs/2020A&A...638A.112V} {638, A112}

\makeatother
\end{thebibliography}



\appendix

\section{Robustness tests}
\subsection{Influence of cluster and infall sample selection on QFE}
\label{appendix:a}
The cluster, infall and control samples of galaxies are selected using photometric redshifts.  Galaxies which are classified as star-forming typically have weaker $4000$\AA\ breaks than the quiescent population, and hence have greater photometric redshift uncertainties.  For the control sample, we expect the redshift intervals immediately below or above the chosen control field interval to contain similar number densities of galaxies and fractions of galaxy type. Therefore we expect a similar level of each type of galaxy is scattered into the control sample as are scattered out. Therefore this effect does not greatly influence the galaxy stellar mass functions or quiescent fractions measured in the control sample. 

For the cluster and infall samples, however, the redshift interval spanning the cluster has a larger galaxy density than the intervals immediately above and below. Furthermore, the cluster and infall contain higher fractions of passive galaxies than the surrounding control field. Hence fewer galaxies are scattered into the cluster/infall compared to the number that are scattered out. This produces a bias in the galaxy SMFs and quiescent fractions measured in the cluster and infall samples relative to the control sample. 

\begin{figure}
\centering
\includegraphics[width=8.0cm]{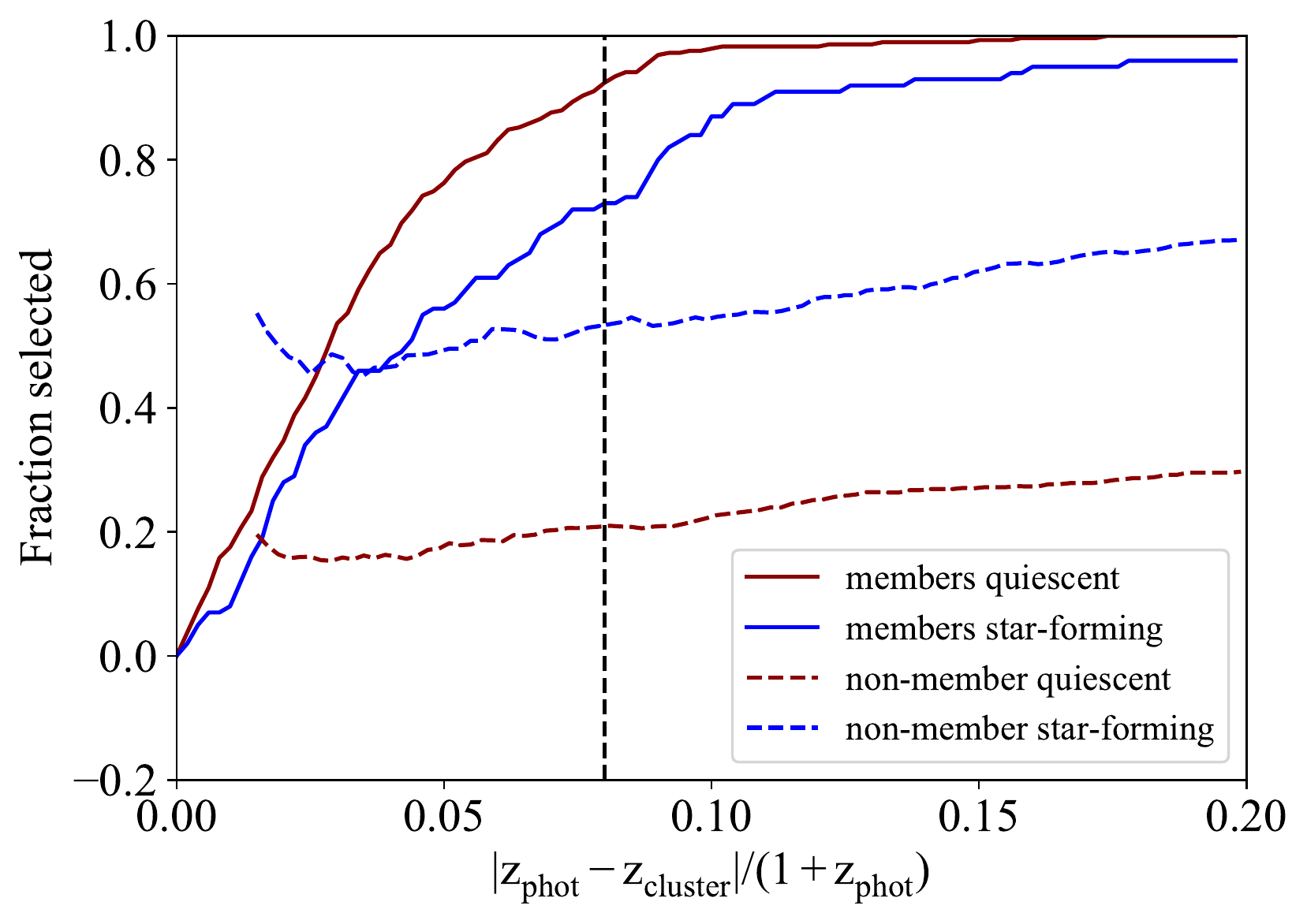}
\caption[]{The completeness (solid lines) and contamination (dashed lines) fraction of the cluster sample selected using a redshift interval of $|z_{\rm phot}- z_{\rm cl}|/(1 + z_{\rm phot})$. The dashed black line marks our fiducial sample selected with $|\Delta  z|/(1 + z)=0.08$. Red and blue lines show quiescent and star-forming subclasses, respectively. Both completeness and contamination rates of the cluster sample increase with $|\Delta z|/(1 + z)$.}
\label{fig:redshift}
\end{figure}

 In Figure \ref{fig:redshift} we demonstrate how the photometric redshift errors affect the completeness (ratio of spectroscopic cluster members selected to all spectroscopic galaxies selected) and contamination (ratio of non-cluster spectroscopic members selected to all spectroscopic galaxies selected) of the cluster sample. We use the cluster membership of the spectroscopic sample as defined in GOGREEN DR1 \citep{Balogh_2021}.
 
 Using the |$\Delta z$|$/(1+z) = 0.08$ interval we select a cluster sample that is greater than $90$\% complete in quiescent galaxies, but only $73$\% complete for star-forming galaxies since a larger fraction of the star forming galaxies have been scattered out of the redshift interval. 
 On the other hand,  $54$\% of the star forming cluster sample are contaminants, whereas only $21$\% of the quiescent galaxies are contaminants. Again, this is because the photometric redshifts of the star forming galaxies are more uncertain so a large fraction of the star forming galaxies in the adjacent intervals are scattered into the cluster sample than quiescent galaxies. 
 
\begin{figure*}
\includegraphics[width=2\columnwidth]{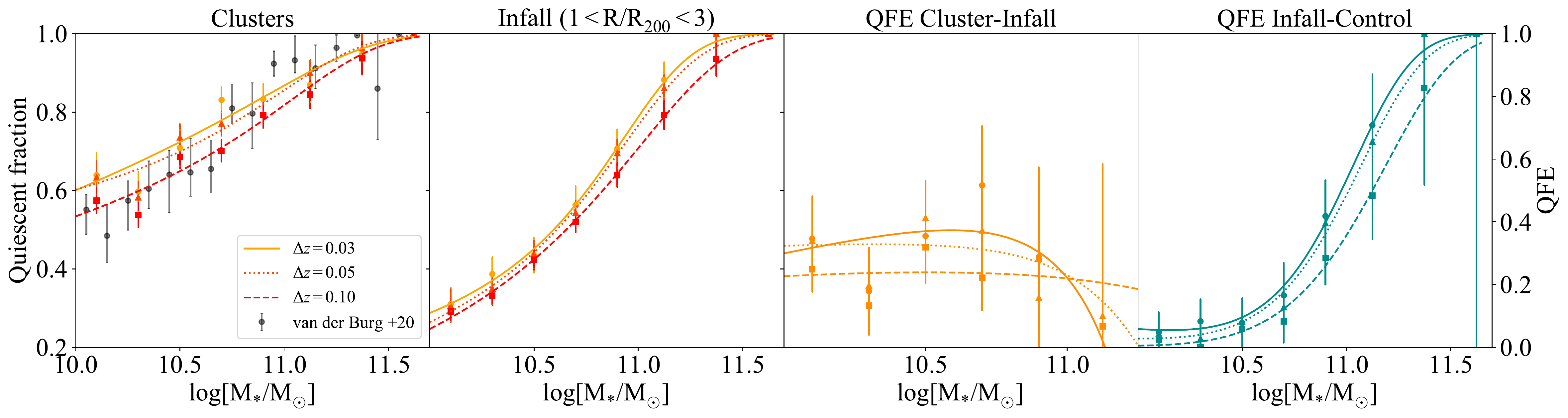}
\caption[]{The quiescent fractions (two left-hand panels) and QFE (two right-hand panels) derived using infall and cluster samples selected with redshift intervals of  $|\Delta z|/(1+z) = 0.03,~0.05,~0.1$. The quiescent fractions increase when using smaller redshift intervals, however the general trends of the QFE$_{cl-pc}$ and  QFE$_{pc-con}$ are consistent with results using the fiducial redshift intervals of $0.08$ in Figures \ref{SMF} and \ref{QFE}. The grey circles in the left-most panel display the quiescent fraction of a subset of GOGREEN clusters measured by \vdBt, which are consistent with the quiescent fractions we measure. }
\label{QFE_deltaz}
\end{figure*}

Figure\,\ref{fig:redshift} further demonstrates how the completeness and contamination fractions increase as the redshift interval is increased. Thus a different redshift interval can affect the quiescent fractions for the cluster and infall galaxies. In the two left panels of Figure\,\ref{QFE_deltaz} we display the quiescent fraction for cluster and infall galaxies selected using the redshift intervals  $|\Delta z|/(1+z) = 0.03,~0.05,~0.1$. For smaller intervals than the fiducial $|\Delta z|/(1+z) = 0.08$ we obtain larger quiescent fractions at all masses. For the higher interval than the fiducial we obtain lower quiescent fractions at all masses. 

We also display the quiescent fraction of the GOGREEN cluster sample measured by \vdBt\ in Figure \,\ref{fig:redshift}. The cluster sample of \vdBt\ is slightly different to our GOGREEN+GCLASS sample, and they make corrections to account for the fraction of quiescent and star-forming galaxies scattered out of the cluster sample using spectroscopic cluster members. Figure \,\ref{fig:redshift} demonstrates that our cluster quiescent fractions are consistent so these corrections are relatively minor. 

In the two right panels of Figure\,\ref{QFE_deltaz} we display the quenched fraction excess, QFE$_{cl-inf}$ and QFE$_{inf-con}$ selected using the redshift intervals $|\Delta z|/(1+z) = 0.03,~0.05,~0.1$. Although the normalisation differs slightly, our results for the fiducial $|\Delta z|/(1+z) = 0.08$ case are consistent within uncertainties. Therefore our conclusions remain valid and are independent of our choice of |$\Delta z$|.

\begin{figure*}

\includegraphics[width=2\columnwidth]{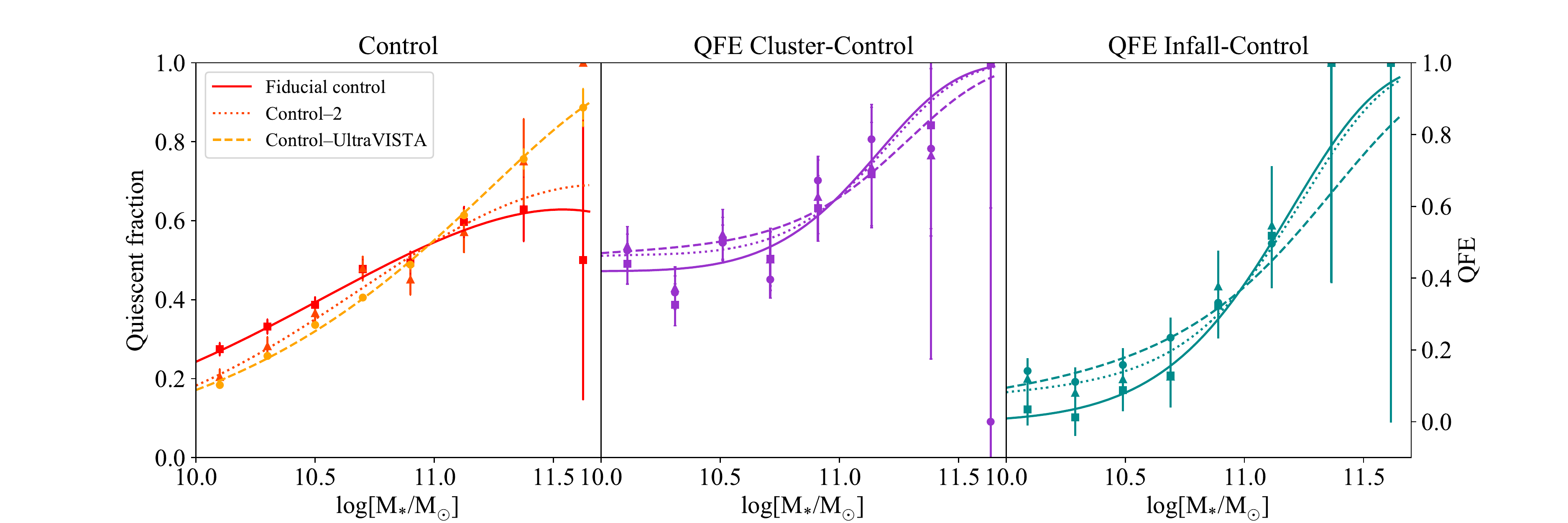}
\caption[]{The quiescent fractions (left), QFE$_{~cl-con}$ (middle) and QFE$_{~inf-con}$ (right) derived using the fiducial, Control--2 and Control--UltraVISTA control galaxy samples. The trends are the consistent with that of the fiducial control field sample with only minor differences in the absolute numbers. We therefore surmise that our conclusions are robust against reasonable variations in the control field selection.}
\label{fig:QFE_fields}
\end{figure*}

\subsection{Influence of control sample selection on QFE}
\label{appendix:b}

We select our control field sample from the same data as the cluster and infall sample in order to account for observational biases inherent in all data. We therefore select our control sample from each of the 15 cluster fields by selecting galaxies that lie at least 1\,Mpc from the cluster centre and in a redshift range of  $0.15<$|$z_{\rm phot}- z_{\rm cl}$|$/(1 + z_{\rm phot})<0.3$. The GOGREEN and GCLASS data of each cluster only span approximately $10\times10$ arcmin$^2$, which means the area for which we can select control galaxies is limited (as shown in Fig\,\ref{RA_DEC}). As a result the control sample is relatively small, consisting of only 2632 galaxies, and is therefore limited by large Poisson noise. Furthermore, due to the necessity of selecting control galaxies at a significantly different redshift range than the infall region, we include galaxies up to $z=2.3$ and apply a larger level of completeness correction to the low mass control field galaxies than the infall and cluster galaxies. 

To investigate how the above mentioned issues may affect our measurements of the QFE we construct two further control samples and recompute the QFEs. Our fiducial control sample, used in the main body of the paper, is constructed from galaxies that lie at least 1\,Mpc from the cluster centre and in a redshift range of  $0.15<$|$z_{\rm phot}- z_{\rm cl}$|$/(1 + z_{\rm phot})<0.3$ within each cluster field. To determine whether the selection of galaxies with redshifts as far as $|\Delta z|\sim0.3/(1 + z_{\rm phot})$ from the cluster and infall samples affects our results we construct a control sample that is limited to the redshift range of our clusters, i.e.\,$0.8<z<1.5$. We select all galaxies within this redshift range that lie beyond 1\,Mpc from the cluster centre. We then remove the infall and cluster members from the sample by removing all galaxies within the redshift limits $|z_{\rm phot}- z_{\rm cl}|/(1 + z_{\rm phot})<0.1$ within each field. This control sample, which we call \lq Control--2,\rq\ consists of 1513 galaxies. Each of these galaxies is associated with a weight, 1/completeness($Ks$), that allows us to correct for incompleteness as done in Section \ref{sec:GSF}.

Both our fiducial control and Control--2 samples consist of relatively few galaxies and so Poisson noise can have a significant impact on our results. To investigate the impact of Poisson noise we construct a further control sample from the UltraVISTA survey by selecting all galaxies with photometric redshifts in the range $0.8<z<1.5$ within the unmasked area of the 1.62\,deg$^2$ DR1 catalogue \citep{Muzzin2013}. This \lq Control--UltraVISTA\rq\ sample consists of 23,687 galaxies with $M>10^{10}$\Msun. We do not apply any completeness corrections since the survey is 95\% complete to our stellar mass limit of $10^{10}$\,\Msun\ up to $z=1.5$ \citep{Muzzin2013}. \vdBt\ note that the rest-frame $U-V$ and $V-J$ colours of the GOGREEN and GCLASS data are offset by 0.05 with respect to UltraVISTA. We therefore apply shifts of $+0.05$ to both the $U-V$ and $V-J$ colours of each galaxy in the UltraVISTA control sample.

We classify galaxies in all control samples as star-forming or quiescent using the criteria in equation\,\ref{eqn:colour} and plot the quiescent fractions in the left panel of Fig.\,\ref{fig:QFE_fields}. Overall, the trends are qualitatively similar but the fraction of quiescent galaxies for higher (lower) masses is greater (smaller) for the Control-2 and UltraVISTA compared to the fiducial Control. 

The Control-UltraVISTA sample contains a high fraction of quiescent massive galaxies relative to the the fiucial control sample we use, although comparible within the Poisson error bars. Although the Poisson noise is smaller for UltraVista, this field contains many galaxy groups that contaminate this control sample, even at $z=1$ \citep{Giodini2012}.
Quiescent galaxies in the highest density regions of such a control field dominates the shape of the SMF, particularly at the massive end \citep{Papovich_2018}. Therefore the quiescent fraction of the highest mass bins in the UltraVISTA control sample contains the largest contamination by galaxy cluster and group members, and therefore is less trustworthy than the small Poisson error bars imply.

In the middle and right panels of Fig.\,\ref{fig:QFE_fields} we show the impact these different control samples have on the  QFE$_{cl-con}$ and QFE$_{inf-con}$, respectively. The QFE trends are the same for all cases with only minor differences in the absolute numbers of the QFE.  Since the use of alternative control samples result in similar QFEs as the fiducial control sample we surmise that Poison noise, cosmic variances and completeness corrections do not greatly affect our conclusions.

\label{lastpage}
\end{document}